\documentclass{article}

\usepackage{arxiv}
\usepackage[utf8]{inputenc} % allow utf-8 input
\usepackage[T1]{fontenc}    % use 8-bit T1 fonts
\usepackage{hyperref}       % hyperlinks
\usepackage{url}            % simple URL typesetting
\usepackage{booktabs}       % professional-quality tables
\usepackage{amsfonts}       % blackboard math symbols
\usepackage{amsmath}       % blackboard math symbols
\usepackage{nicefrac}       % compact symbols for 1/2, etc.
\usepackage{microtype}      % microtypography
\usepackage{lipsum}		% Can be removed after putting your text content
\usepackage{graphicx}
\usepackage[numbers]{natbib}
\usepackage{doi}
\usepackage{import}
\usepackage{lineno}
\usepackage{caption}
\usepackage{subcaption} 
\usepackage{tikz}
\usepackage{tabularray}
\usepackage{array}
\usepackage{amssymb}
\usepackage{authblk} % for authors info and affiliations
\usepackage[inline]{enumitem}

\def\metagen{\textit{MetaGen}}

\usepackage{xspace} % add whitespace after word only if needed

\newcommand{\suppnamejournal}{appendix\xspace} % lowercase

\title{Inverse Design of Diffractive Metasurfaces Using Diffusion Models}

\author[1, *]{\textbf{Liav Hen}}
\author[1]{\textbf{Erez Yosef}}
\author[1]{\textbf{Dan Raviv}}
\author[1]{\textbf{Raja Giryes}}
\author[1,2, $\dagger$]{\textbf{Jacob Scheuer}}

\affil[1]{School of Electrical and Computer Engineering, Tel-Aviv University, Israel}
\affil[2]{The Center for Nanosciences and Nanotechnology, Tel-Aviv University, Israel}
\affil[*]{email: \texttt{liavhen@gmail.com}}
\affil[$\dagger$]{email: \texttt{kobys@tauex.tau.ac.il}}

\date{} % Empty date to remove date display

\renewcommand{\headeright}{}
\renewcommand{\undertitle}{}
\renewcommand{\shorttitle}{}

\hypersetup{
    pdftitle={Inverse Design of Diffractive Metasurfaces Using Diffusion Models},
    pdfauthor={Liav Hen, Erez Yosef, Dan Raviv, Raja Giryes, Jacob Scheuer},
    pdfsubject={Physics - Optics (physics.optics), Machine Learning (cs.LG)},
    pdfkeywords={Metasurfaces, Inverse Design, Diffusion Models, Nanophotonics, Computational Optics, Inverse Problems},
    pdfcreator={LaTeX with hyperref package},
    colorlinks=true,
    linkcolor=blue,
    citecolor=blue,
    urlcolor=blue
}

\begin{document}
\maketitle

\begin{abstract}
Metasurfaces are ultra-thin optical elements composed of engineered sub-wavelength structures that enable precise control of light. Their inverse design - determining a geometry that yields a desired optical response - is challenging due to the complex, nonlinear relationship between structure and optical properties. This often requires expert tuning, is prone to local minima, and involves significant computational overhead.
In this work, we address these challenges by integrating the generative capabilities of diffusion models into computational design workflows. Using an RCWA simulator, we generate training data consisting of metasurface geometries and their corresponding far-field scattering patterns. We then train a conditional diffusion model to predict meta-atom geometry and height from a target spatial power distribution at a specified wavelength, sampled from a continuous supported band. Once trained, the model can generate metasurfaces with low error, either directly using RCWA-guided posterior sampling or by serving as an initializer for traditional optimization methods. We demonstrate our approach on the design of a spatially uniform intensity splitter and a polarization beam splitter, both produced with low error in under 30 minutes. To support further research in data-driven metasurface design, we publicly release our code and datasets.\footnote{\url{https://github.com/liavhen/metagen}}
\end{abstract}

% keywords can be removed


\section{Introduction}
\begin{figure}[h!]
  \centering
  \begin{minipage}[c]{0.47\textwidth}
    \centering
    \begin{subfigure}[c]{0.52\textwidth}
      \centering
      \includegraphics[width=\linewidth]{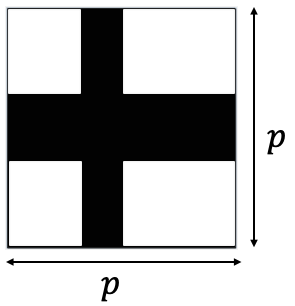}
      \caption{A meta-atom represented as a binary image.}
      \label{fig:binary-meta-stom}
    \end{subfigure}
    \vspace{1em}
    \begin{subfigure}[c]{0.55\textwidth}
      \centering
      \includegraphics[width=\linewidth]{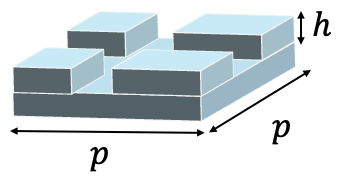}
      \caption{A schematic of a meta-atom.}
      \label{fig:3d-meta-atom}
    \end{subfigure}
  \end{minipage}%
  \hfill
  \begin{minipage}[c]{0.52\textwidth}
    \centering
    \begin{subfigure}[c]{\textwidth}
      \centering
      \includegraphics[width=\linewidth]{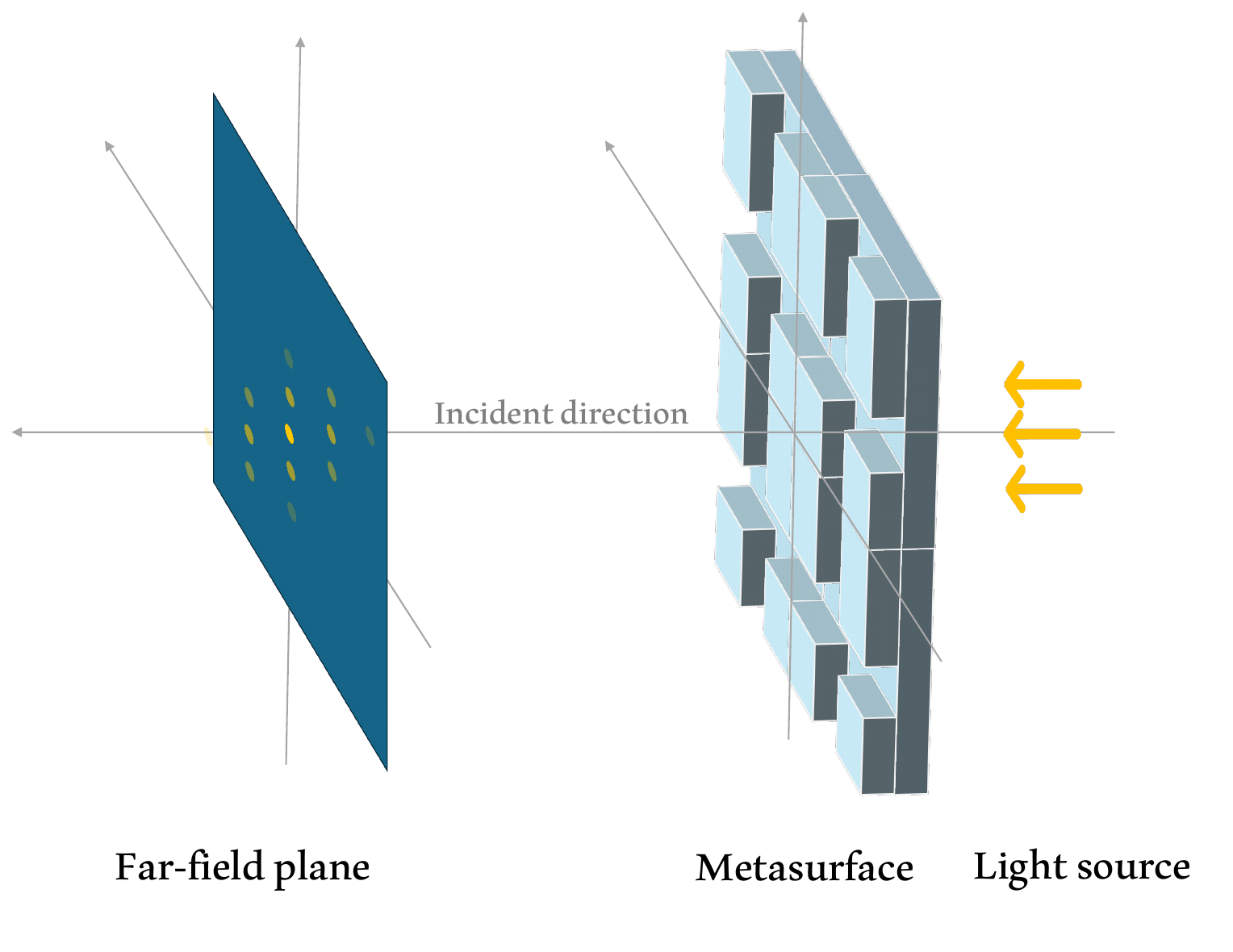}
      \caption{A schematic of the scattering scenario: A periodic metasurface, excited by a monochromatic plane-wave.}
    \label{fig:3d-whole-cell}
    \end{subfigure}
  \end{minipage}
  \vspace{-2em}
\end{figure}

Diffractive metasurfaces are optical devices composed of sub-wavelength meta-atoms that enable precise control over light propagation. Designing these structures to achieve specific optical properties remains a significant challenge due to the complex and non-linear relationship between the geometry and the optical response.

Early research on metasurface inverse design focused on classical optimization techniques. In particular, gradient-based methods relying on numerical light-scattering simulations, such as Rigorous Coupled-Wave Analysis (RCWA)~\cite{RCWAMoharam95}, were extensively explored~\cite{torcwa, Peurifoy2018, Kim2020diffraction}. However, these approaches often converge to local minima. To address this, alternative strategies, such as topology optimization, evolutionary algorithms, and particle swarm optimization, have been proposed~\cite{TopologyOptimization4InverseDesign, EvolutionaryAlgorithms4InverseDesign}. These are frequently combined with global phase initialization schemes, such as the Iterative Fourier Transform Algorithm (IFTA), to improve convergence~\cite{hao2019diffraction, Li2021diffraction, Yan2023Diffraction}. While effective, these methods are computationally intensive, often requiring hours to design a single metasurface, and depend on expert-tuned regularization strategies.

With the rise of deep learning, neural networks (NNs) have also been explored for metasurface design~\cite{Peurifoy2018, tanriover2023generative, Diffusion4InverseDesign1, Diffusion4InverseDesign2, Diffusion4InverseDesign3}. However, these models often underperform analytical methods due to their reliance on large, task-specific datasets which are rarely available in practice.

It is important to note that most prior work on metasurface inverse design using NNs has focused on the spectral response as the primary design objective. In contrast, controlling the far-field spatial power distribution, namely the directionality and intensity of scattered Bragg orders, has received limited attention, despite its importance in beam splitting, shaping, and multiplexing applications.

This paper addresses the inverse design of metasurfaces that produce a desired spatial far-field power distribution. We consider periodic, binary-thickness dielectric metasurfaces under normal incidence. Since the wavelength and geometric features are of comparable scale, the periodicity induces scattering into discrete diffraction orders. Our goal is to recover a meta-atom configuration that yields a prescribed angular power distribution at specified wavelengths.

To this end, we have developed a generative framework based on diffusion models~\cite{SongDiffusion, ddim} - a class of neural generative algorithms increasingly used to generate high-fidelity samples and solve inverse problems. We train the model on a large dataset of simulated structure–response pairs, constructed using a novel, scalable data generation pipeline. It should be emphasized that in contrast to the more common tasks tackled by NNs, such as image classification, generation or text prediction, a large training dataset does not exist and must be constructed ad hoc. Our data generation method addresses the lack of high-quality task-specific data and can be adapted to various nanophotonic design problems.

Once trained, diffusion models can be sampled to generate solutions that match the learned distribution. We explore two operational modes:
\begin{itemize}
    \item \textbf{RCWA guidance (posterior sampling):} full-wave simulations (e.g., RCWA) are integrated into the diffusion generative process, trading off accuracy vs. runtime.
    \item \textbf{Integrative mode:} The diffusion model provides initial guesses for downstream optimization routines.
\end{itemize}

It should be noted that the training process (especially the dataset generation method - see Section~\ref{sec:method}) creates a distribution of topology and diffraction pattern pairs. It can be expected that some of the metasurface topologies (and, hence, the attainable diffraction patterns) are not well-represented within this distribution. Consequently, the design of metasurfaces targeting such diffraction patterns is a challenging task for our diffusion model. We tackle this gap through incorporation of careful considerations to the data generation process and state-of-the-art methods for querying diffusion models in light or desired measurements consistency. 

We demonstrate the effectiveness of our approach on several challenging design tasks, particularly targeting patterns which are not well-represented within the distribution of the learned dataset. These include the generation of SiO$_2$ metasurfaces exhibiting an X-shaped scattering pattern across multiple wavelengths, the design of a silicon dual-polarization uniform beam splitter, and a polarization splitter refined via gradient descent, all within minutes.

This paper is organized as follows: Section~\ref{sec:method} details the data generation process, the inverse design process, and evaluation methodology. Section~\ref{sec:results} presents experimental results and demonstrates the usage modes of our diffusion model for metasurface inverse design.

%%%%%%%%%%%%%%%%%%%%%%%%%%%%%%%%%%%%%%%%%%%%%%%%%%%%%%%

\section{Method}\label{sec:method}

In this section, we present our proposed method for the inverse design of diffractive metasurfaces using diffusion models. We begin by introducing a novel data generation procedure. We then describe the training of our diffusion model, which we designate as \metagen, and how it is used to design new metasurfaces. Finally, we outline our evaluation methodology, which includes comparisons with alternative generative frameworks and the design of metasurfaces corresponding to target scattering patterns. Additional details are provided in Sections~\ref{app:data}-\ref{app:implementation-details} of the \suppnamejournal.

\subsection{Data Generation}  \label{sec:data}

\begin{table}[ht]
\centering
\renewcommand{\arraystretch}{1.3} % Default value: 1
\begin{tabular}{ccc}
\hline
Dataset                                 & B2                            & C2                    \\
\hline
Polarizations                           & TE Only                       & TE \& TM              \\
Material                                & SiO$_2$ ($n \approx 1.45$)    & Si ($n \approx 3.6$)  \\
Wavelengths $\lambda$ $[\mu m]$         & $\{0.8, 0.85, 0.9, 0.95, 1.0\}$ & $\{1.4, 1.45, 1.5, 1.55, 1.6\}$\\
Heights $h$ $[\mu m]$                   & $0.75-1.25$                 & $0.1-0.6$           \\
Periodicity  $p$ $[\mu m]$              & $2.86$                         & $4.6$                 \\ 
\hline
\end{tabular}
\vspace{1em}
\caption{\textbf{Physical settings of the main datasets.}  
The B2 and C2 datasets were generated using our method and are used for training. Both are designed to ensure the presence of scattering patterns with at least \(5 \times 5\) propagating Bragg orders. Additional datasets and configurations are described in the \suppnamejournal.}
\label{tab:datasets}
\end{table}

The datasets consist of pairs of binary images that represent the topology of the metasurface and the transmitted and reflected scattering efficiencies $T$ and $R$, which are represented as matrices. Each pixel in these matrices contains the total power projected into one of the Bragg diffraction orders, where the central pixel corresponds to the zero-order diffraction lobe.

\paragraph{Binary Structures Randomized Generation.} To construct a diverse dataset of binary metasurface structures, we developed a custom generation pipeline that produces meta-atoms from randomized noise. The process begins by applying a low-pass spatial filter to realizations of uniformly sampled noise $\sim \mathbb{U}(0, 1)$. While the random sampling promotes diversity and free-form variation, the low-pass filter enforces a minimum feature size, ensuring compatibility with standard fabrication constraints. We perform the spatial filtering in the frequency domain using a 2D Fourier transform, which naturally enforces periodic boundary conditions in the generated meta-atoms. The filtered noise is then binarized, where each resulting binary pattern encodes a metasurface design: \texttt{1} denotes dielectric material, and \texttt{0} denotes air. To further diversify the dataset, we generate additional parameterized shapes, such as rectangles, ellipsoids and grating profiles - structures which may not naturally arise from free-form generation. For implementation details, see Section~\ref{app:data} of the \suppnamejournal.

\begin{figure}[h!]
    \centering
    \includegraphics[width=1\linewidth]{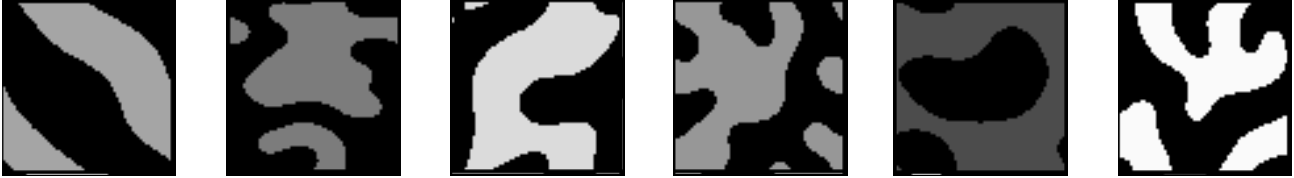}
    \caption{\textbf{Binary Structures from our Proposed Dataset}. Different gray levels represent different thicknesses $h$. By design, the structures are naturally periodic and continuous.}
    \label{fig:structures}
\end{figure}

\paragraph{Simulations.} The light scattering patterns from the metasurfaces are computed using Rigorous Coupled-Wave Analysis (RCWA)~\cite{RCWAMoharam95}, a numerical method for solving Maxwell’s equations in periodic structures by representing electromagnetic fields as truncated Fourier series. We use a GPU-accelerated, differentiable implementation of RCWA~\cite{torcwa}. 
In each dataset, the metasurface thickness and operating wavelength vary uniformly across samples, while other physical parameters, such as the periodicity and material composition, are fixed. All simulations are carried out under normal incidence, though this assumption can be relaxed by generating datasets corresponding to oblique illumination conditions.

\paragraph{Datasets.} To demonstrate our method under varying physical conditions, we constructed two main datasets, B2 and C2. Both datasets are designed to exhibit at least 5x5 propagating orders in air, but differ in the material, the supported wavelengths and the periodicity. Additional datasets with a different number of excited orders are detailed in Section~\ref{app:data} in the \suppnamejournal. Both datasets consist of 3.6M (720K data points per operating wavelength), with the resolution of the binary images set to 64x64 pixels.

\subsection{Inverse Model}

\begin{figure}[h]
    \centering
    \includegraphics[width=1\linewidth]{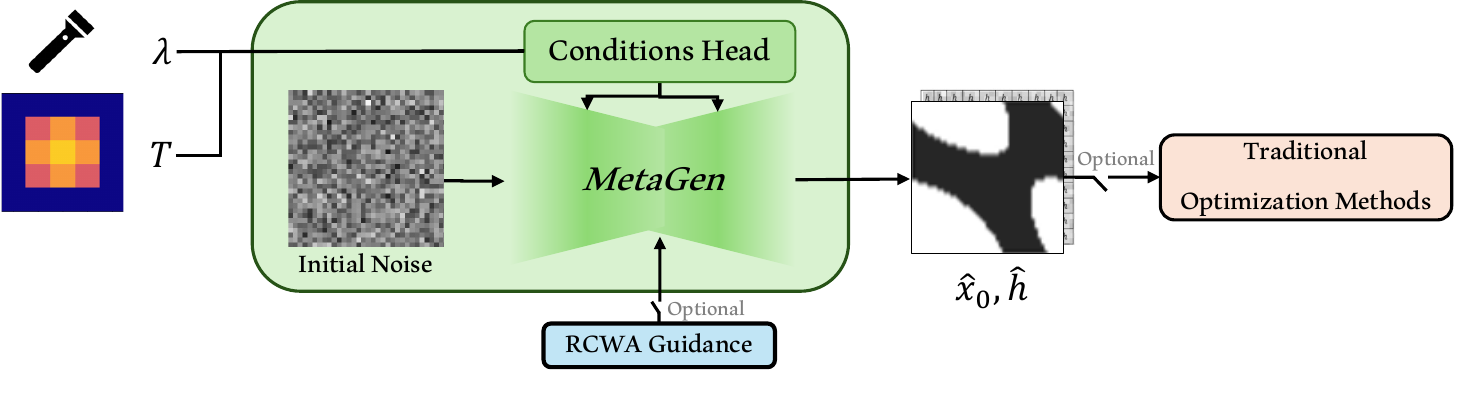}
    \caption{\textbf{\metagen: Diffusion-based inverse design.}  
    We train a diffusion model to reconstruct the geometry of a metasurface, specifically, the binary pattern \( x \) and height \( h \), starting from pure noise and conditioned on the target far-field scattering pattern and operating wavelength. The model's output is denoted by \( \hat{x}_0, \hat{h} \), following the conventional notation used in diffusion models.}
    \label{fig:metagen-scheme}
\end{figure}

We train a diffusion model, \metagen, to serve as an inverse predictor, reconstructing the geometrical attributes of a metasurface - specifically, its binary structure and height - given a target spatial power distribution and an operating wavelength \( \lambda \). The model outputs a 2D binary image representing the meta-atom layout, along with a corresponding scalar height value.

Training is performed separately on the B2 and C2 datasets (see Table~\ref{tab:datasets}) using a standard diffusion modeling approach. Samples from the training set are corrupted with additive Gaussian noise, and the model is trained to denoise them while conditioned on the corresponding target scattering patterns and wavelengths. Through this process, the model learns the statistical relationship between metasurface geometry and its optical response. The training procedure follows the Elucidated Diffusion Models (EDM) framework~\cite{EDM}, with further details provided in Sections~\ref{app:background} and~\ref{app:implementation-details} of the \suppnamejournal.

\paragraph{Sampling.}  
Once trained, \metagen\ can be sampled through an iterative process to generate valid metasurface designs starting from pure noise. See Figure~\ref{fig:metagen-scheme} for a schematic overview. To further enforce consistency with physical measurements, we employ \emph{Diffusion Posterior Sampling} (DPS)~\cite{chung2022diffusion}. This technique integrates gradients from a forward simulator (RCWA in our case) into the diffusion trajectory, guiding the generation toward samples that are both physically plausible and consistent with the target scattering pattern. In our context, we refer to this approach as \emph{RCWA guidance}. 
This technique requires backpropagating through the simulator, which increases computational cost in exchange for improved accuracy. In practice, we apply RCWA guidance either at all diffusion steps or only at selected ones, depending on the complexity of the target design task. We note that the model can also be sampled directly without RCWA guidance, enabling near-instant generation. While this unguided approach performs well on target patterns drawn from the training distribution, we observe reduced effectiveness when generalizing to idealized or out-of-distribution target diffraction patterns.

\subsection{Evaluation}
We evaluate model performance by comparing the target scattering patterns \( T, R \), which condition the generation process along with the wavelength \( \lambda \), to the actual scattering patterns \( \hat{T}, \hat{R} \) obtained from simulating the predicted meta-atom structures using the RCWA simulator. Given its relevance in many applications, our evaluation primarily focuses on the relative error in transmittance (\( T \)). When assessing performance across multiple components (e.g., two polarizations), the relative error is computed for each component separately and then averaged.

\paragraph{Metrics.} We propose using relative error as the metric, as it intuitively represents the relative amount of power that is either missing or incorrectly distributed between the diffraction lobes compared to the desired scattering pattern. The error metric is defined as follows for $\textbf{x} \in \{T, R\}$:

\begin{equation}\label{eq:relative_error}
    \mathrm{e}(\textbf{x}, \hat{\textbf{x}}) = \frac{\left\| \textbf{x} - \hat{\textbf{x}} \right\|_1}{\left\| \textbf{x}  \right\|_1}
\end{equation}

\paragraph{Evaluation Strategy.} Since \metagen\ relies on the data distribution used during training, it is important to evaluate its performance not only on the test set, drawn from the same distribution as the train set, but also on manually constructed scattering patterns representative of real-world applications. We generally refer to such scattering patterns as \emph{target patterns}. In Section~\ref{sec:results}, we present three such application-driven designs obtained using our approach. To further assess the effectiveness of diffusion models in capturing the training distribution, and also generalizing to target patterns beyond the observed distribution, we compare \metagen\ with other generative baselines. Specifically, we evaluate a modified Wasserstein Generative Adversarial Network (GAN) with Gradient Penalty (WGAN-GP)~\cite{arjovsky2017wasserstein, gulrajani2017improved, mirza2014conditional, Cohen2023GAN}, adapted for conditional sampling, and a Conditional Variational Autoencoder (CVAE)~\cite{sohn2015learning, Bank2023AE} designed for similar inverse tasks. A detailed comparison is provided in Section~\ref{app:evaluation-details} of the \suppnamejournal.

\paragraph{Visualization.} Scattering patterns are shown as heat maps, with each Bragg order $(m_x, m_y)$ color-coded by intensity. Values above 0.01 are reported with two decimal places.

%%%%%%%%%%%%%%%%%%%%%%%%%%%%%%%%%%%%%%%%%%%%%%%%%%%%%%%%%%%%%%%%%%%%%%%%%%%%%%%%%%

%%%%%%%%%%%%%%%%%%%%%%%%%%%%%%%%%%%%%%%%%%%%
\section{Results} \label{sec:results} 

This section presents results demonstrating the effectiveness of \metagen\ in solving real-world optical design tasks. First, we present its broad spectral support by generating metasurfaces that produce scattering patterns that are beyond the learned distribution across multiple wavelengths. More specifically, we show the ability of \metagen\ to design a metasurface exhibiting X-shaped scattering patterns. Next, we showcase the design of a spatially uniform scattering pattern, achieving performance comparable to prior works but with significantly reduced runtime. Finally, we demonstrate the use of \metagen\ to generate candidate metasurface designs that serve as initializations for auxiliary optimization procedures, thus reducing significantly the design time. As a concrete example we demonstrate fast design of a metasurface based polarization beam splitter.

As illustrated in Figure~\ref{fig:metagen-scheme}, the sampling process begins from a random noise realization, which gradually transforms into a fresh sample valid under the training distribution, i.e., a metasurface. Thus, the generation process is essentially stochastic, and multiple candidate metasurfaces may be designed by initializing the sampling process with different noises. We leverage this variability, together with the parallelism capabilities of modern GPU-accelerated frameworks, to improve design outcomes with only marginal runtime overhead. For each design task presented in this study, we generate \(k\) candidate samples in parallel and select the one that yields the lowest scattering error.

\subsection{Wide Spectral Support} \label{subsec:direct-sampling}

\begin{figure}[h]
    \centering
    \includegraphics[width=1\linewidth]{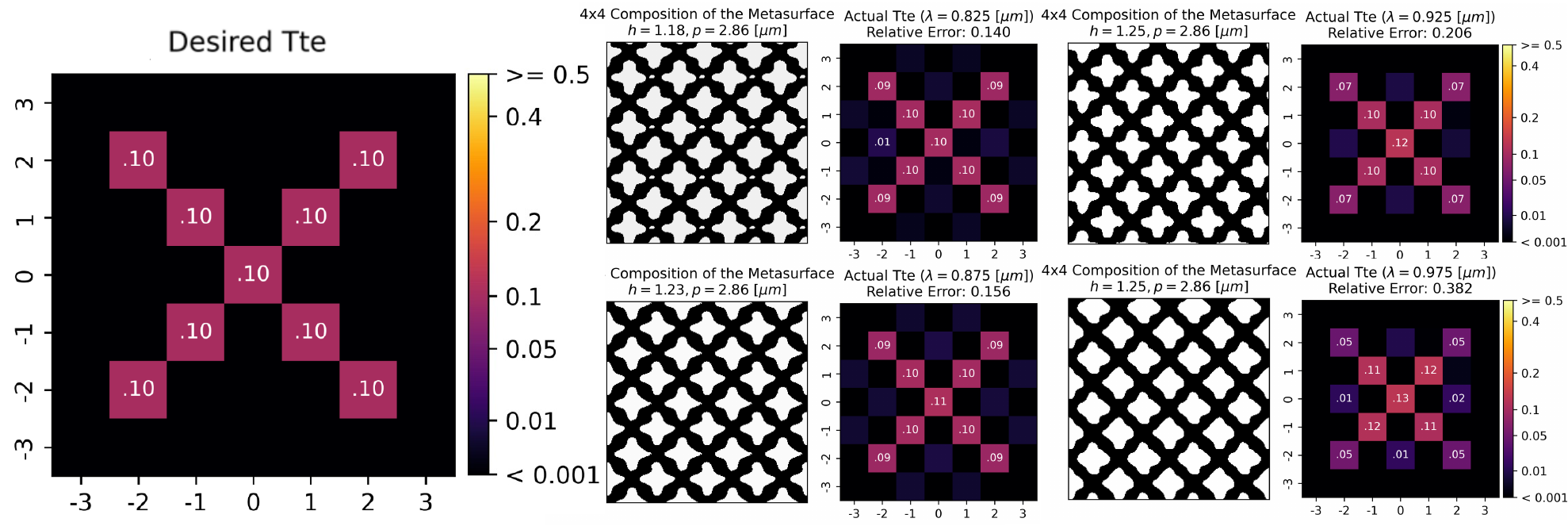}
    \caption{\textbf{Wide spectral support.}  
    Our model is trained on multiple discrete wavelengths, enabling it to generalize to unseen wavelengths through interpolation. In this example, we design metasurfaces that produce an X-shaped scattering pattern across several operating wavelengths outside the training distribution.}
    \label{fig:x-shape}
\end{figure}

We use \metagen\ to generate a non-trivial (out of the learned distribution) X-shaped scattering pattern across a range of wavelengths: $ \{0.825, 0.875, 0.925, 0.975\}~[\mu\text{m}] $. Note that none of these wavelengths is present in the training set, demonstrating the model's broad spectral support and its ability to interpolate between trained wavelengths. This design is conducted under the B2 configuration, i.e., SiO$_2$ based metasurface with TE-polarized incident light only (see Table~\ref{tab:datasets}). Figure~\ref{fig:x-shape} shows the resulting metasurface designs (for each wavelength) and the actual scattering patterns they produce.

\subsection{Spatially Uniform Intensity Beam Splitter}

In this section, we demonstrate the design of a beam splitter, i.e., a metasurface that uniformly distributes the intensity of the incident plane wave across a 5$\times$5 Bragg diffraction orders. Prior studies~\cite{hao2019diffraction, Kim2020diffraction, Li2021diffraction, Yan2023Diffraction} have addressed this task using optimization procedures that typically require several hours. Our goal is to significantly accelerate the design process while maintaining comparable performance. In contrast with prior works, we choose to showcase a design that is simultaneously constrained to exhibit a uniform scattering pattern in both polarizations. This represents an extension of the original challenge, which typically considers only a single polarization mode. Following these works, we evaluate our design using the Uniformity Error (UE) metric:

\begin{equation} \label{eq:ue}
    \mathrm{e}_\text{UE}(\hat{T}) = \frac{T_\text{max} - T_\text{min}}{T_\text{max} + T_\text{min}}, \quad \text{where} \quad T_\text{max} = \max_i T_i, \quad T_\text{min} = \min_i T_i,
\end{equation}

where \( \{T_i\} \) denotes the transmitted power in each diffraction order.

\begin{figure}[h]
    \centering
    \includegraphics[width=1\linewidth]{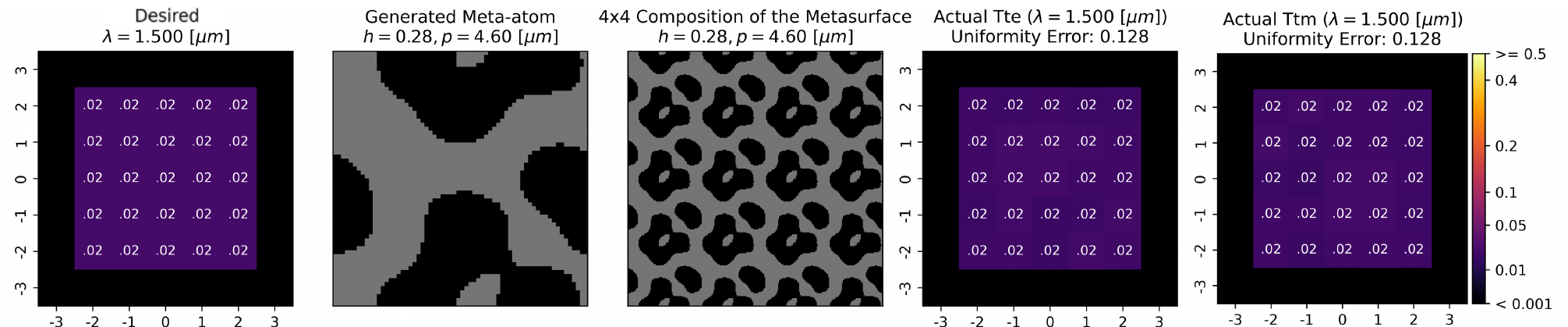}
    \caption{\textbf{Uniform Beam-Splitting Metasurface.}  Our approach integrates information from RCWA measurements directly into the diffusion sampling process. This allows us to overcome local-minima sensitivity without expert tuning and to achieve high-quality results within minutes. In this example, we present a uniform, dual-polarization 5\(\times\)5 beam-splitting silicon metasurface designed for normally incident light at wavelength \( \lambda = 1.5~\mu\text{m} \).}
    \label{fig:dual-polarized-uniform}
    \vspace{-1em}
\end{figure}

The results are summarized in Table~\ref{tab:ue-comparison}. Our model, trained under the C2 configuration, achieves comparable performance to those of previous works while simultaneously constraining both polarizations. The corresponding design is shown in Figure~\ref{fig:dual-polarized-uniform}.  Notably, our solution was generated in just 28 minutes using a best-of-\(k = 30\) sampling strategy, in contrast to the hours-long runtimes reported in previous works. To achieve this, we use 200 diffusion steps and apply RCWA-based guidance during the final 100 steps.

\begin{table}[h]
    \centering
    \begin{tabular}{ccccc}
        \hline
        Method & Hao et al.~\cite{hao2019diffraction} & Li et al.~\cite{Li2021diffraction} & Yan et al.~\cite{Yan2023Diffraction} & \metagen\ (Ours) \\
        \hline
        UE (Eq.~\ref{eq:ue}) & 0.1740 & 0.1264 & 0.1176 & 0.1280 \\
        Runtime & 3.5 hours\footnotemark[1] & 2.6 hours & 18.9 hours & 28 minutes \\
        \hline
    \end{tabular}
    \vspace{1em}
    \caption{\textbf{Inverse design of a 5$\times$5 uniform beam splitter.} 
    Compared to prior works, our method achieves comparable UE while completing in a fraction of the runtime.\protect\footnotemark[2]}
    \label{tab:ue-comparison}
    \vspace{-1em}
\end{table}

\footnotetext[1]{The runtime was estimated based on the reported duration of 90 seconds per generation and visual analysis of figures showing approximately 140 generations.}
\footnotetext[2]{Runtime comparisons may be affected by differences in hardware. See Section~\ref{app:hardware} of the \suppnamejournal\ for full hardware specifications.}

\subsection{Polarization Splitter: Integrating Traditional Optimization}

\begin{figure}[h]   
    \centering
    \includegraphics[width=1\linewidth]{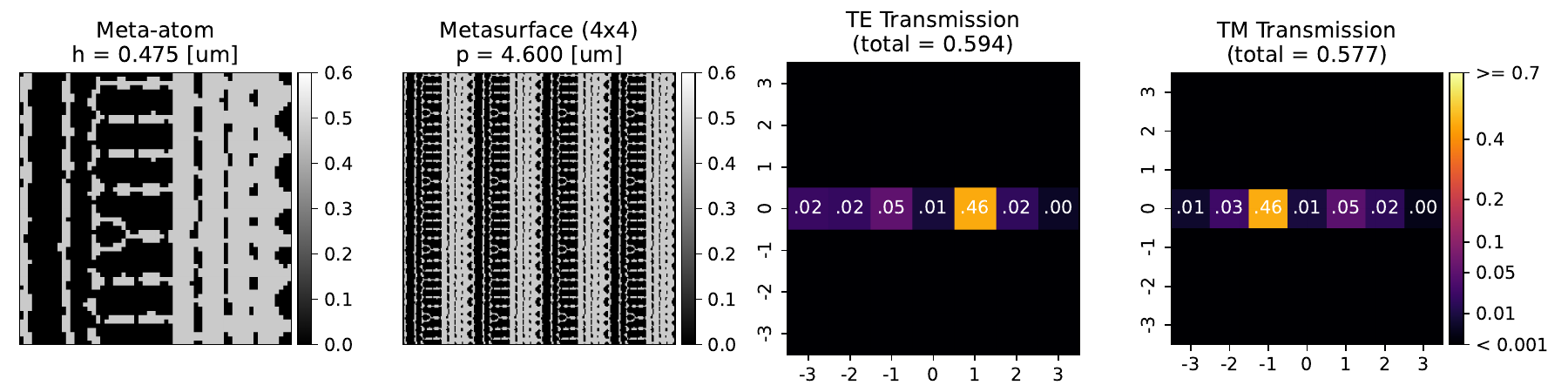}
    \caption{\textbf{Polarization splitter design using \metagen.} Candidate patterns generated by \metagen\ were used as initializations for gradient descent optimization, consistently outperforming randomly initialized baselines. Here, 800 optimization steps were performed with a decaying learning rate, and the objective function was the Mean Squared Error (MSE). Each pixel contains the fractional amount of transmitted power.}
    \label{fig:beam-splitter}
    \vspace{-2em}
\end{figure}

Recent studies on the inverse design of diffractive metasurfaces commonly employ conventional optimization techniques, initialized randomly or using phase-response methods such as IFTA. Given the vast design space and susceptibility to local minima, population-based algorithms are frequently used to improve convergence ~\cite{hao2019diffraction, Yan2023Diffraction}. While effective, these approaches are typically computationally intensive and require expert-tuned initialization through IFTA.

We propose leveraging \metagen\ as an expert-free, data-driven initializer within such optimization pipelines. By providing good initial guesses, \metagen\ mitigates local-minima convergence issues and significantly accelerates the overall design process. As a reference method, we adopt Gradient Descent (GD) with smoothing and gradual projection steps, as outlined in~\cite{Kim2020diffraction, torcwa}. The smoothing step involves applying a spatial low-pass filter to suppress high-frequency features, while the gradual projection step progressively enforces binarization. Further details on the GD-based implementation for metasurfaces inverse design are provided in Section~\ref{app:gd-details} of the \suppnamejournal.

In this example, we target the design of a \emph{polarization beam splitter}, a device that directs TE- and TM-polarized incident waves into different diffraction orders. We begin by querying \metagen, which was trained on dual-polarized Silicon dataset C2 (see Table \ref{tab:datasets}), to generate patterns that aim to split both polarizations evenly between the \( (1,0) \) and \( (-1,0) \) diffraction orders. These patterns are then used as an initialization for GD, which is applied to refine the design and enhance the scattering response.

\begin{figure}[h]
    \centering 
    \includegraphics[width=1\linewidth]{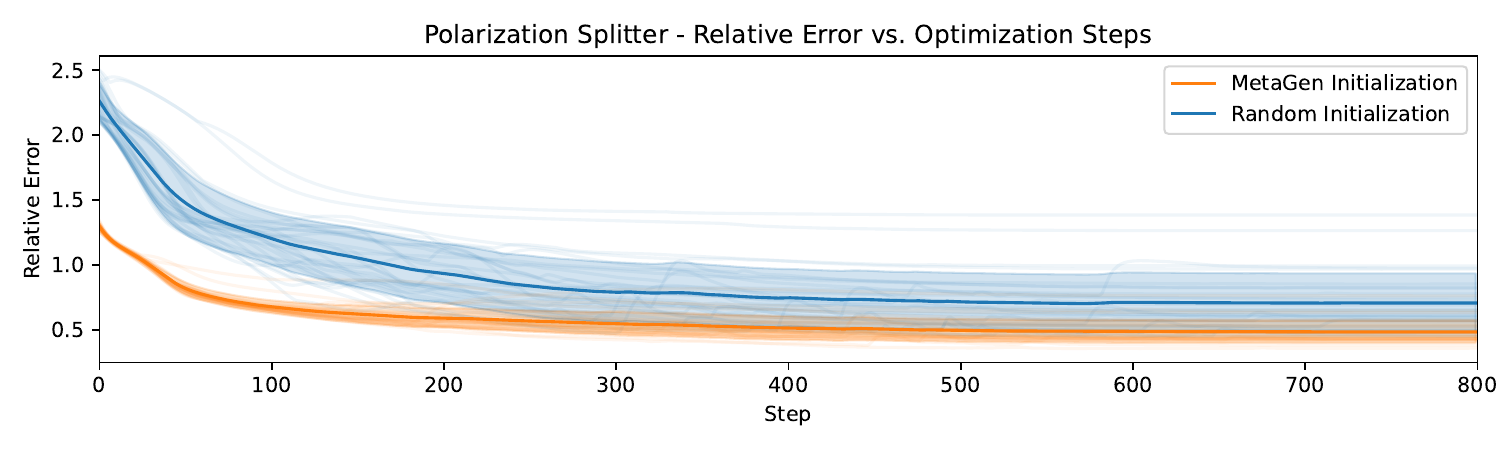}
    \caption{\textbf{Optimization performance for polarization splitting.} The blue and orange curves show the average relative error across 30 randomly and model-initialized runs, respectively. Shaded regions denote standard deviation.}
    \label{fig:optimization-p-splitter}
    \vspace{-1em}
\end{figure}

Figure~\ref{fig:beam-splitter} presents an example of a polarization splitter generated using this hybrid approach. Our results show that initializing auxiliary optimization methods such as GD with outputs from \metagen\ consistently improves final performance, as illustrated in Figure~\ref{fig:optimization-p-splitter}. Moreover, the reduced variance of model-initialized optimization trajectories could also benefit population-based methods, such as genetic algorithms and particle swarm optimization. 

In the hybrid approach, sampling from \metagen\ is performed directly, i.e., without applying RCWA guidance, thus completes almost instantly. The subsequent optimization process requires approximately 8 minutes on the hardware used in this study. Overall, high-quality metasurface designs can be generated within a matter of minutes. Further details regarding the computational setup are provided in Section~\ref{app:hardware} of the \suppnamejournal.

\subsection{Additional Results and Baseline Comparisons}
For quantitative evaluation, we compare our diffusion model with other generative approaches, which we also train on data sets constructed using our data generation method. We provide WGAN-GP and C-VAE baselines on additional datasets (C1–C3) of increasing structural complexity. As detailed in Supplementary Section 3, \metagen\ consistently achieves lower relative error on both test and target patterns (Table S2, Fig. S6), demonstrating superior accuracy and generalization beyond the training distribution. Visual results further support its advantage in modeling complex structure–response mappings (Figs. S4–S7). Notably, these findings align with prior works, in which diffusion models were consistently proven to outperform GANs also in other applications ~\cite{Dhariwal2021DiffusionMB, DiffusionBeatGANTO}.

%%%%%%%%%%%%%%%%%%%%%%%%%%%%%%%%%%%%%%%%%%%%%%%%%%%%%%%

%%%%%%%%%%%%%%%%%%%%%%%%%%%%%%%%%%%%%%
\section{Conclusion} \label{sec:conclusion}
In this work, we introduce a diffusion model-based framework for the inverse design of metasurfaces conditioned on specified spatial scattering patterns across multiple incident wavelengths. Our method can operate in a standalone manner to generate metasurfaces with low error. Additionally, accuracy can be improved at the cost of increased computation time by integrating a RCWA simulator into the sampling process, enabling posterior refinement. Furthermore, recognizing the limitations of traditional optimization methods, which often require expert-tuned initialization or incur long runtimes, we demonstrate how our model can serve as a fast, expert-free initializer to accelerate and improve the performance of such methods.
Our diffusion model is trained on paired data consisting of metasurface structures and their corresponding scattering patterns. To support this approach, we develop a novel, scalable data generation pipeline capable of producing training data at any scale, for a wide range of physical settings and tasks. We construct two primary datasets encompassing diverse physical conditions to enable systematic evaluation of inverse design methods. By designing applicable optical devices, such as a uniform beam splitter and a polarization beam splitter, we show that our diffusion model achieves performance comparable to state-of-the-art approaches that typically require hours of optimization, while significantly reducing computational overhead. These results highlight the potential of diffusion models for scalable and efficient metasurface design, opening avenues for further research in data-driven nanophotonics.

%%%%%%%%%%%%%%%%%%%%%%%%%%%%%%%%%%%%%%%%%%%%%%%%%%%%%%%%%%%%%%%%%%%%%%%%%%%%%%%%%%
\section*{Acknowledgements}
We thank Doron Klepach and Guy Ben-Dov from FVMat and Shady Abu-Hussein for fruitful discussions.
% \end{acknowledgement}
%%%%%%%%%%%%%%%%%%%%%%%%%%%%%%%%%%%%%%%%%%%%%%%%%%%%%%%%%%%%%%%%%%%%%%%%%%%%%%%%%%

% \bibliographystyle{unsrt}
% \bibliography{main-references}

%%%%%%%%%%%%%%%%%%%%%%%%%%%%%%%%%%%%%%

\clearpage
\section*{Appendix}
\appendix

\section{Background} \label{app:background}

In this section, we briefly introduce the main concepts that form the foundation of this study.

\subsection{Diffusion Models}\label{sec:diffusion-models}

Diffusion models are a class of generative models that synthesize data by reversing a gradual noising process. Both the noising and denoising processes, also termed the forward and reverse processes, can be modeled using stochastic differential equations (SDEs)~\cite{SongDiffusion}. The forward process adds noise to clean data, while the generative task is formulated as solving the corresponding reverse-time SDE to recover structured samples from noise.

Let \( x(t) \in \mathbb{R}^d \) denote a sample evolving under a forward SDE of the form:
\begin{equation}\label{eq:forward-sde}
    \mathrm{d}x = f(x, t)\,\mathrm{d}t + g(t)\,\mathrm{d}w,
\end{equation}
where \( f(x, t) \) is a drift term, \( g(t) \) is a time-dependent noise scale, and \( w \) is standard Brownian motion. This forward diffusion gradually corrupts the data, yielding a tractable prior distribution, typically standard Gaussian, at the final time \( t = T \).

The reverse-time SDE, used for sampling, is given by~\cite{anderson1982reverse}:
\begin{equation}\label{eq:reverse-sde}
    \mathrm{d}x = \left[f(x, t) - g(t)^2 \nabla_x \log p_t(x) \right]\,\mathrm{d}t + g(t)\,\mathrm{d}\bar{w},
\end{equation}
where \( \nabla_x \log p_t(x) \) is the \emph{score function}, the gradient of the log-density of the noised data at time \( t \), and \( \bar{w} \) denotes reverse-time Brownian motion. Alternatively, one can sample from the \emph{probability flow ODE}, which shares the same marginals but evolves deterministically~\cite{SongDiffusion}:
\begin{equation}\label{eq:prob-flow}
    \mathrm{d}x = f(x, t)\,\mathrm{d}t - \frac{1}{2}g(t)^2 \nabla_x \log p_t(x)\,\mathrm{d}t.
\end{equation}

To generate samples from the data distribution \( p_{\text{data}}(x) \), one must solve either the reverse SDE or the probability flow ODE (Eq. \ref{eq:reverse-sde} and \ref{eq:prob-flow}). However, both require access to the score function, which is intractable in practice, but can be computed using \emph{Tweedie's Formula} \cite{efron2011tweedie}:

\begin{equation}\label{eq:tweedie}
    \mathbb{E}[x_0 | x_t] = x_t + \sigma^2(t) \nabla_x\log p_t(x)
\end{equation}

where $x_t, x_0$ abbreviate $x(t), x(0)$ and $\sigma(t)$ is the noise level at time $t$, depending on the choice of $g(t)$ in Eq. \ref{eq:forward-sde}. Previous works~\cite{hyvarinen2005estimation, vincent2011connection, SongDiffusion, EDM} have shown that $\mathbb{E}[x_0 | x_t]$ can be approximated using a neural network $D_\theta(x_t, t)$ minimizing the expected $\mathcal{L}_2$ denoising loss. Namely, if for every time $t$, $D_\theta(x_t, t)$ minimizes $\mathbb{E}_{x_0 \sim p_{\text{data}}, x_t \sim p(x_t|x_0)} \left[ \left\| D_\theta(x_t, t) - x_0 \right\|_2^2 \right]$, where $p(x_t|x_0)$ follows the forward SDE (Eq. \ref{eq:forward-sde}), then:

\begin{equation}\label{eq:approx-score-function}
    s_\theta(x_t, t) := \frac{1}{\sigma^2(t)}\left ( D_\theta(x_t,t) - x_t\right ) \approx \nabla_x \log p_t(x)  
\end{equation}

With the approximated score function from Eq. \ref{eq:approx-score-function}, new samples can be generated by discretizing and solving the reverse-time SDE or the probability flow ODE (Eq. \ref{eq:reverse-sde} and \ref{eq:prob-flow}) using numerical solvers. In this work, we follow the elucidated formulation of~\cite{EDM}.

\subsection{Inverse Problems}

Inverse problems are typically formulated as:
\begin{equation}
    y = \mathcal{A}(x) + \eta,
\end{equation}
where \( y \) is a measurement of the unknown signal \( x \), obtained via a forward  operator \( \mathcal{A} \), and \( \eta \) represents measurement noise. Recovering $x$ from $y$ is often difficult because either \( \dim(y) \ll \dim(x) \) or \( \mathcal{A} \) is non-injective, making the problem ill-posed.

In this work, we cast the metasurface inverse design task as a noiseless (\( \eta \equiv 0 \)) inverse problem, where \( x \) denotes the metasurface geometry and height, \( y \) is the diffraction pattern, and \( \mathcal{A} \) is the measurement operator implemented using a RCWA simulation~\cite{RCWAMoharam95, torcwa}. For generality, we use conventional inverse problem notation throughout this section.

\subsection{Solving Inverse Problems with Diffusion Models}

When the operator \( \mathcal{A} \) is non-linear or non-convex, solving the optimization problem
\begin{equation}
    \arg\min_x \left\| y - \mathcal{A}(x) \right\|_2    
\end{equation}

can be challenging in practice. Alternatively, especially due to the possible multiplicity of solutions, one can take a Bayesian approach and solve:
\begin{equation}
    \arg\max_x \, p(x \mid y) = \arg\max_x \, p(x)p(y \mid x) = \arg\max_x \, \log p(x) + \log p(y \mid x),
\end{equation}
where the second equality follows from Bayes’ theorem and the last from taking the logarithm. This splits the objective into a prior term \( \log p(x) \) and a likelihood term \( \log p(y \mid x) \).

To solve such problems without direct access to the densities \( p(x), p(y|x) \), one can leverage diffusion models by modifying the score-based sampling equations (Eq.~\ref{eq:reverse-sde} or Eq.~\ref{eq:prob-flow}), replacing the unconditional score with a conditional one, as follows:
\begin{equation}
    \nabla_x \log p_t(x \mid y) = \nabla_x \log p_t(x) + \nabla_x \log p_t(y \mid x) \approx s_\theta(x_t, t) + \nabla_x \log p_t(y \mid x).
\end{equation}

The prior term is approximated by the trained denoiser \( \epsilon_\theta \), but the likelihood term remains intractable in the general case and requires further derivations. Two main strategies have been explored to address this.

\paragraph{Conditional Diffusion Models.} This approach involves training the denoising network \( D_\theta(x_t, t, y) \) to explicitly condition on \( y \), with the aim of directly learning the conditional score, i.e., \( \nabla_x \log p_t(x \mid y) \approx s_\theta(x_t, t, y) \). Notably, ~\cite{ho2022classifier} show that by randomly hiding the conditional input $(y = \varnothing)$, one can train a conditional diffusion model $D_\theta(x_t, t, y)$ and an unconditional one $D_\theta(x_t, t, \varnothing)$ simultaneously. It is further shown that by extrapolating the conditional and unconditional scores with a scale $w$, an enhanced sampling can be achieved, termed \emph{classifier-free guidance}.:

\begin{equation}
    s^{(w)}_\theta(x_t, t, y) =  s_\theta(x_t, t, \varnothing) + w (s_\theta(x_t, t, y) - s_\theta(x_t, t, \varnothing))
\end{equation}

\paragraph{Guided Diffusion Models.} In contrast, guidance-based methods do not require conditional training. Instead, they approximate the likelihood gradient using a differentiable forward operator \( \mathcal{A} \), assuming access to it. In deep learning frameworks, the gradients of such operators are available due to the backpropagation algorithm. In the Gaussian case ($\eta \sim \mathcal{N}$), ~\cite{chung2022diffusion} suggest to approximate the likelihood gradient as:
\begin{equation}
    \nabla_x \log p_t(y \mid x) \approx \gamma_t \nabla_x\left\| y - \mathcal{A}(\hat{x}_0(x_t)) \right\|^2_2,
\end{equation}

where 

\begin{equation}
    \hat{x}_0(x_t) = x_t + \sigma^2(t)s_\theta(x_t, t) \left ( = D_\theta(x_t, t) \right )
\end{equation}

is the Tweedie estimator as in Eq. \ref{eq:tweedie}, and \( \gamma_t \) is a scaling factor. This technique, known as Diffusion Posterior Sampling (DPS), results in a guided conditional score:
\begin{equation}
    \nabla_x \log p_t(x \mid y) \approx s_\theta(x_t, t) + \gamma_t \nabla_x\left\| y - \mathcal{A}(\hat{x}_0(x_t)) \right\|^2_2.
\end{equation}

In this work, we combine both approaches. Following \cite{ho2022classifier}, we train an unconditional and a conditional diffusion model together on paired data samples \( (x, y) \). While conditional sampling alone often produces good results, it was observed that combining it with posterior guidance improves both reconstruction accuracy and robustness, effectively enables the generation of samples beyond the training distribution. We note that although our inverse design problem for diffractive metasurfaces involves no measurement noise, we still adopt the DPS formulation for the Gaussian case, due to its observed positive impact on performance.

\section{Data}\label{app:data}

\subsection{Data Generation Details}\label{app:data-generation-details}

The data generation process involves filtering uniform noise realizations in the frequency domain using a 2-D Fourier Transform, which ensures computational efficiency and cyclic-induced operations that produce binary images with natural periodic continuity. Then, the smoothed noise is binarized to obtain a binary defined structure. In the frequency domain, a circularly symmetrical low-pass filter is applied to the spatial frequency magnitude $r = \sqrt{\nu_x^2 + \nu_y^2}$, where $\nu_x, \nu_y$ are the spatial frequencies of the image. The low-pass filter is characterized by $r_c$, the cut-off radial frequency, and $\beta$, a smoothing factor which we set to 3, as defined in Eq. \ref{eq:lpf}. To align with fabrication limitations, the cut-off frequency was set to $r_c = \frac{1} {l_\text{min}}$, with $l_\text{min}$ the minimal feature size allowed in the spatial domain. However, while performing spatial smoothing suppresses the existence of high-frequency features, projecting it either $0$ or $1$ re-introduces high-frequencies. We observe that iteratively smoothing and projection mitigates this phenomena, and promotes better enforcement of the minimal feature size criteria.

\begin{equation}\label{eq:lpf}
    H(r) = \frac{1}{4} \left(1 + \tanh{(\beta(r + r_c))}\right)\left(1 + \tanh{(-\beta(r - r_c)})\right)
\end{equation}

\paragraph{Choosing the generation parameters.}  
We focus on metasurfaces composed of square-shaped unit cells. For such structures, under normal incidence, a diffraction order \( (m_x, m_y) \) can propagate in air only if it satisfies the condition \( m_x^2 + m_y^2 \leq \left(\frac{p}{\lambda}\right)^2
 \)~\cite{Loewen2018DiffractionGA}. As a result, each dataset is characterized by a specific number of propagating diffraction orders. We define the height and minimal feature size of the meta-atoms given the choice of material and fabrication constraints. The operating wavelengths are selected to encourage high transmission efficiency. Based on these wavelengths, the periodicity \( p \) is then set to ensure that the resulting metasurfaces support the desired number of propagating orders.

\begin{figure}[h!]
    \centering
    \includegraphics[width=1\linewidth]{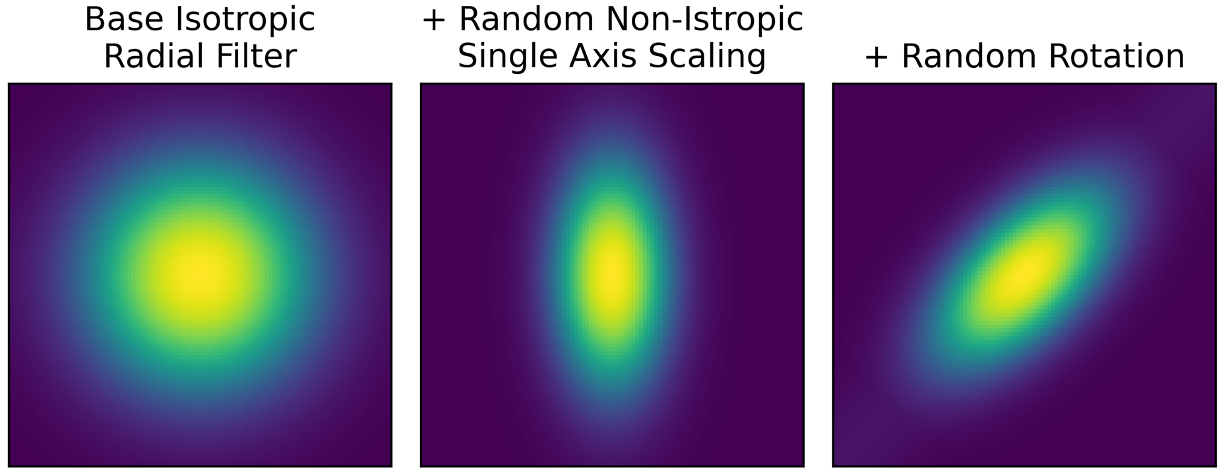}
    \caption{\small\textbf{Visualization of $H(r)$ with Random Variations.} Each sample is generated by filtering standard uniform noise using a 2D radial filter, as shown here, with occasional axis scaling and rotation applied to the filter in Fourier space to encourage diversity in the output data points.}
    \label{fig:h-filter}
\end{figure}

\paragraph{Introducing structural biases.} During this study, we observe that, under absolute structural freedom, meta-atoms frequently exhibit somewhat similar scattering patterns. In practice, many interesting applications, such as prisms, beam-splitters and more, require some inherent structural order such as symmetry and directionality. Addressing this observation, we introduce additional steps for biasing the random generation procedure towards such structural properties. First, after generating a random noise realization $n$, we randomly draw $\gamma \in \{0, 0.1, 0.2, 0.3, 0.4, 0.5\}$, and modify the noise to be $\hat{n} = \gamma n + (1-\gamma)\tilde{n}$, where $\tilde{n}$ is a horizontally flipped version of $n$. This step introduces pseudo-symmetry, with $\gamma=0.5$ enforcing actual horizontal symmetry of the meta-atom and $\gamma=0$ corresponding to an absolute free-form structure. We then rotate the image in a random angle to encourage the existence of symmetry in all directions. Next, we randomize $l_\text{min}$, allowing for structures with larger characteristic features. Then, within the application of the low-pass filter, we randomly shrink one of the axes $\nu_x$ or $\nu_y$, and then rotate the filter, resulting in a non-isotropic filter with respect to the periodic axes. Figure \ref{fig:h-filter} visualizes these heuristics.

\paragraph{Combining other parameterized distributions.} While our biased free-form generation method covers a broad range of scattering patterns, we augment the dataset with additional parameterized families of structures that are unlikely to be produced through free-form generation alone:
\begin{enumerate*}
    \item Rectangles.
    \item Ellipsoids.
    \item Jerusalem crosses.
    \item Pseudo-freeform structures (from a publicly available collection of reciprocal designs\footnote{\url{https://github.com/SensongAn/Meta-atoms-data-sharing}}~\cite{pseudo-freeform-dataset}).
    \item Grating profiles.
\end{enumerate*}

Figure~\ref{fig:structures} shows representative examples from each family. For the pseudo-freeform and parameterized families (i.e., families 1–4), additional variation was introduced by occasionally compressing structures along one axis, forming artificial sub-periods.

In total, 80\% of the dataset was generated using the free-form method, and 4\% was allocated to each of the families 1–5.

\begin{figure}[h!]
    \centering
    \includegraphics[width=1\linewidth]{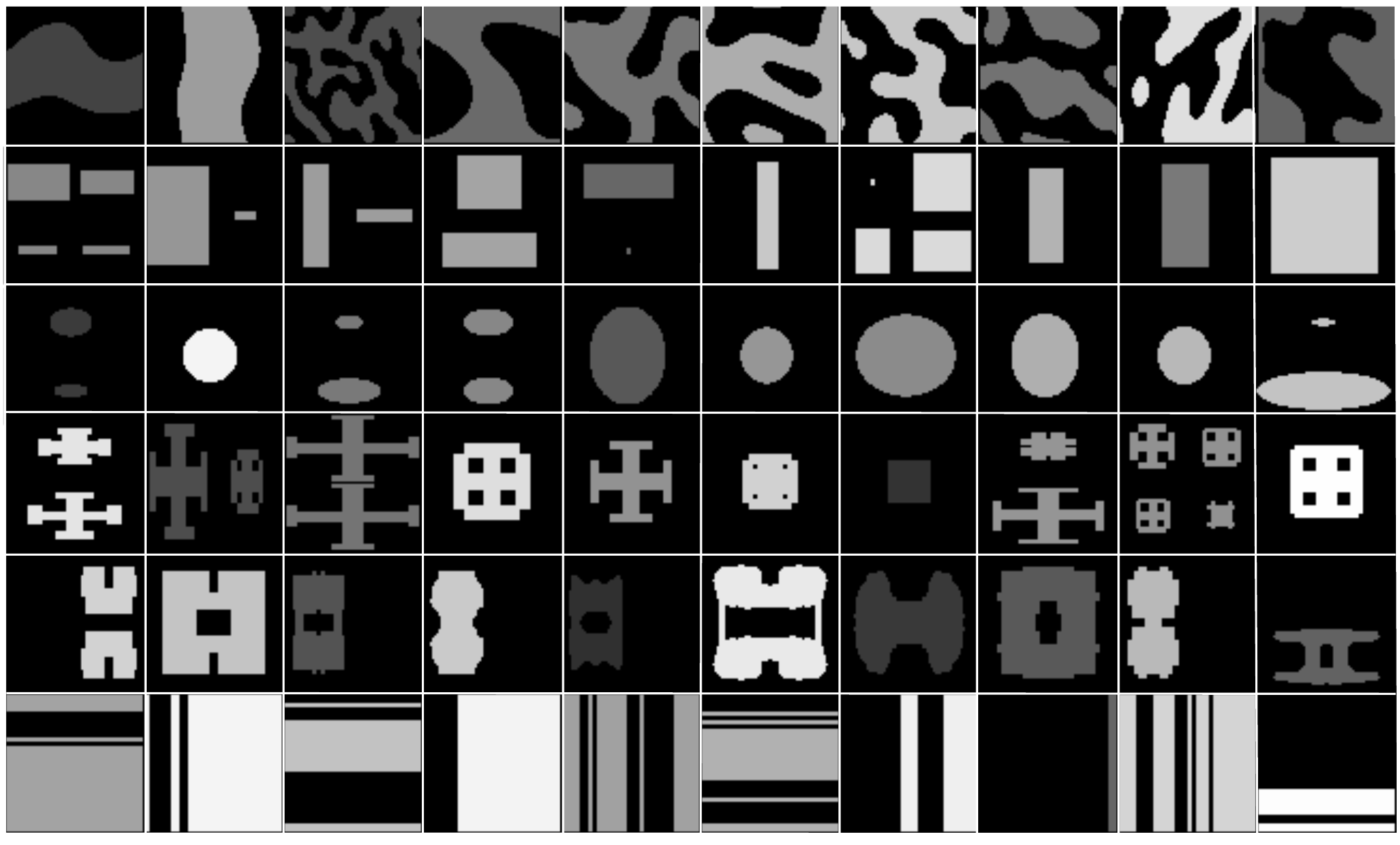}
    \caption{\small\textbf{Examples of meta-atom structures}. Each row shows representative samples from a different family.}
    \label{fig:structures}
\end{figure}

\paragraph{Accuracy.} The RCWA algorithm exploits the periodicity of metasurfaces by representing their structure using a truncated Fourier expansion~\cite{RCWAMoharam95}. This formulation enables a trade-off between computational efficiency and simulation accuracy, controlled by the number of retained Fourier orders.  In our experiments, we prioritize speed while maintaining acceptable accuracy by using 7 Fourier orders for SiO$_2$ datasets and 9 orders for Si datasets. The higher setting for Si reflects its larger refractive index contrast with air, which requires finer resolution to capture sharp field variations. We validate these choices by calculated total transmitted and reflected energy, allowing numerical errors within a 5\% margin.

\subsection{Analysis of The Computed Scattering Distribution}\label{app:scattering-distribution-analysis}
The RCWA simulations, conducted across various thicknesses and incident wavelengths, map the geometric design space onto a new, unknown distribution of diffraction patterns. The randomization method used to generate the metasurface structures significantly influences this distribution, potentially creating dominant modes or underrepresented regions. Since diffusion models are designed for approximating the training data distribution, ensuring sufficient coverage of the scattering space is crucial for enabling diverse and representative sampling of scattering patterns. To evaluate this coverage, we propose a qualitative method that leverages Principal Component Analysis (PCA) for dimensionality reduction and Kernel Density Estimation (KDE) for visualization of the distribution’s density, allowing examination of the resulted distribution.

PCA is a commonly used technique for linear dimensionality reduction. It identifies the principal directions, or principal components, along which the data shows the most variance. In our case, the target scattering patterns are processed by computing their covariance matrix, from which the $N$ largest eigenvectors are extracted to define these directions of maximum variance. Each data point is then projected onto the principal plane, spanned by the selected $N$ principal components. This plane minimizes the mean Euclidean distance between the data points and their projections onto it, allowing high-dimensional data to be visualized in two dimensions when $N=2$ while highlighting the data’s primary variances.
Once the scattering patterns are projected onto a 2D plane, we apply Kernel Density Estimation (KDE) to smooth the empirically obtained data points and approximate the continuous density of the underlying distribution. KDE achieves this by placing a kernel (e.g., Gaussian) at each data point and summing their contributions across the space. This provides a smoothed density map, helping visualize the distribution of scattering patterns derived from the randomized metasurface structures, as demonstrated in Figure \ref{fig:scattering_distribution_pca}. In this figure, the density of the computed scattering distribution is empirically estimated and visualized. Selected target patterns are then projected onto the resulting principal plane, enabling navigation within the reduced space and validating the coverage of these patterns.\\

\noindent
\begin{minipage}[c]{0.55\textwidth}
To summarize, the coverage of the scattering space can be assessed through the following steps:
\begin{itemize} 
\item Sample a large set of scattering patterns. 
\item Perform PCA with $N=2$ and project the scattering patterns onto the attained plane. 
\item Visualize the 2D plane using KDE for a smoothed density representation. 
\item Construct a set of target patterns of interest, including potentially unachievable ones, and project them onto the plane. 
\end{itemize}
\end{minipage}%
\hfill
\begin{minipage}[c]{0.4\textwidth}
  \centering
    \includegraphics[width=\textwidth]{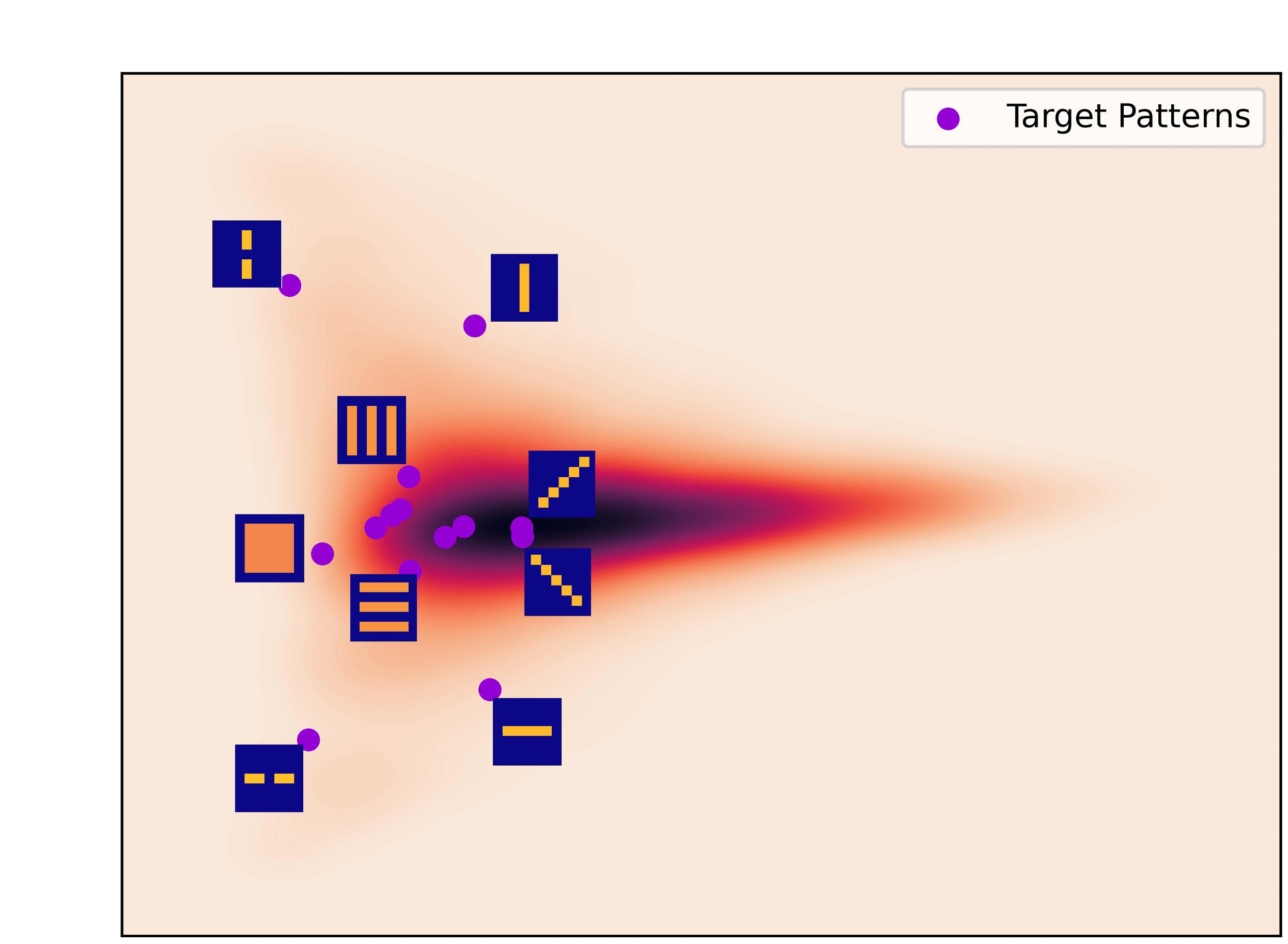}
    \captionof{figure}{\textbf{Scattering Space Visualization.} Target patterns projected on the 2-D principal plane of 5000 data points drawn from the training distribution of dataset A2.}
    \label{fig:scattering_distribution_pca}
\end{minipage}

\hfill \break
We propose this visualization as macro-level inspection of the distribution. Ideally, a fully covered scattering space would be desirable for robust training of distribution leaning algorithms such as diffusion models. However, in our study, achieving this was found to be challenging, and we leave further exploration of this task for future research.

\subsection{Additional Datasets}

In addition to datasets B2 and C2, we construct three additional SiO$_2$-based datasets (A1, A2, and A3) with larger periodicity values, as detailed in Table~\ref{tab:supp-datasets}. These datasets include only free-form structures generated using our method and are designed to empirically test how well our approach scales with growing complexity, in comparison to alternative methods.

\begin{table}[ht]
\centering
\renewcommand{\arraystretch}{1.3}
\begin{tabular}{cccc}
\hline
Polarization & Material & Wavelengths \( \lambda \) [\textmu m] & Heights \( h \) [\textmu m] \\

TE only      & SiO$_2$  & \( \{0.85, 0.9, 0.95, 1.0, 1.05, 1.1\} \) & \( 0.75\,\text{--}\,1.45 \) \\
\hline
\multicolumn{1}{c}{Datasets} & \multicolumn{2}{c}{Periodicity \( p \) [\textmu m]} & \multicolumn{1}{c}{Data Size} \\
\multicolumn{1}{c}{A1 / A2 / A3} & \multicolumn{2}{c}{1.9 / 3.2 / 4.6} & \multicolumn{1}{c}{1.5M (250k$\times$6)} \\
\hline
\end{tabular}
\vspace{1em}
\caption{\small\textbf{Physical settings of the datasets A1, A2, and A3.} These datasets differ only in the periodicity of the meta-atoms, enabling a controlled evaluation of how increasing structural complexity impacts the performance of the proposed method.}
\label{tab:supp-datasets}
\end{table}

\section{Evaluation Details and Quantitative Results}\label{app:evaluation-details}

\subsection{Comparison to Competing Methods}  
To the best of our knowledge, this is the first study to perform distribution learning for metasurface generation conditioned on diffractive scattering patterns. For consistent evaluation, we train all methods - \metagen, WGAN-GP, and C-VAE - on the same datasets.

We compare the models across three key aspects: (i) their ability to capture the training distribution, (ii) generalization to a reference set of target patterns, and (iii) robustness to increasing complexity. To this end, all methods are trained on datasets A1, A2, and A3.

The evaluation is conducted by querying each model with a desired scattering pattern. The models generate corresponding meta-atom structures, which are then simulated using RCWA to obtain their actual scattering patterns. The resulting patterns are compared to the target patterns using relative error as the evaluation metric. Further architectural and training details for each method are provided Section ~\ref{app:implementation-details} in this supplementary material.

\subsection{Evaluation on Data Distribution} \label{sec:evaluation-data-dist}
\begin{figure}[h!]
    \centering
    \begin{subfigure}{\textwidth}
         \centering
         \includegraphics[width=1\textwidth]{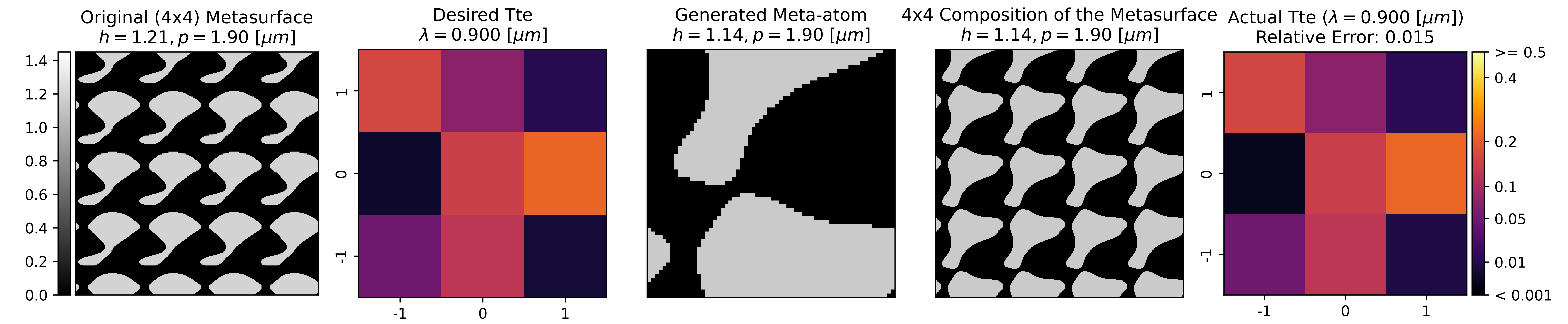}
    \end{subfigure}
    \begin{subfigure}{\textwidth}
         \centering
         \includegraphics[width=1\textwidth]{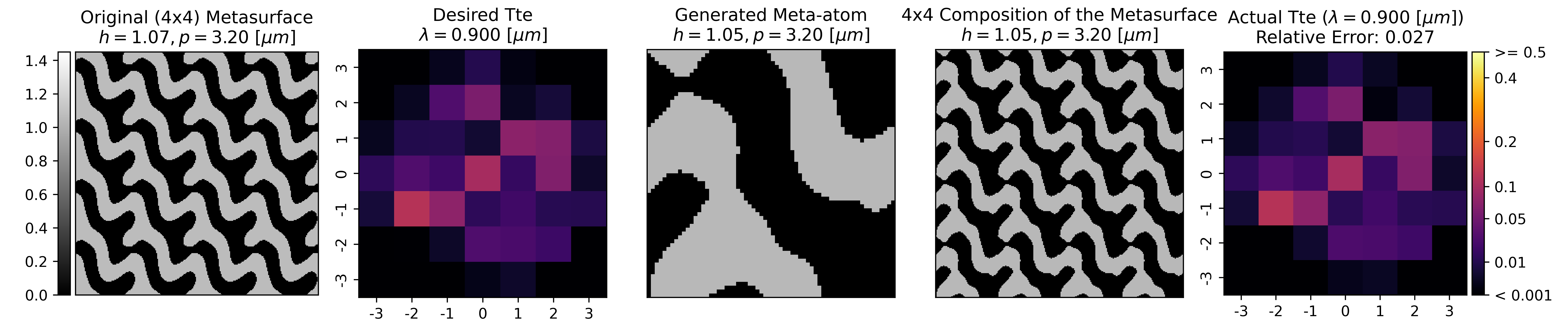}
    \end{subfigure}
    \begin{subfigure}{\textwidth}
         \centering
         \includegraphics[width=1\textwidth]{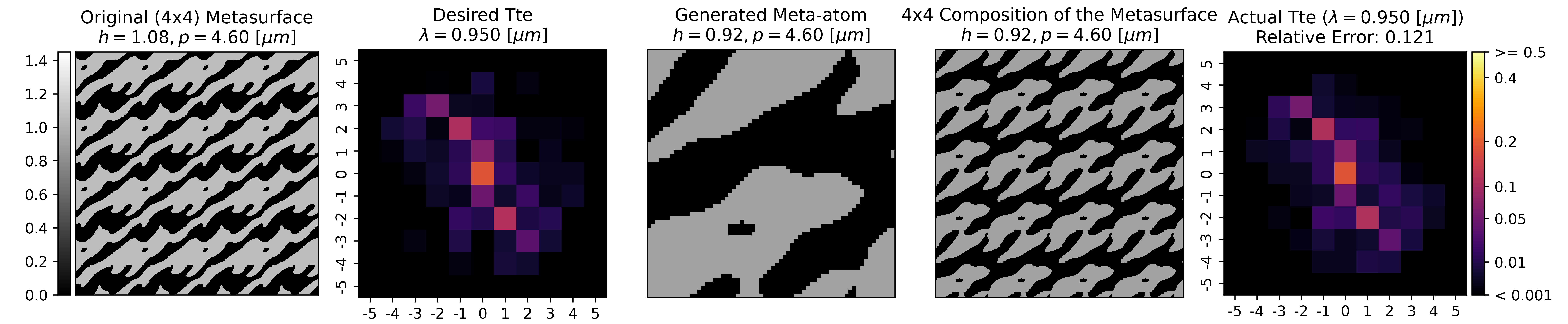}
     \end{subfigure}
    \caption{\small\textbf{Representative In-Distribution Samples}. Metasurfaces designed by sampling \metagen\ conditioned on scattering patterns drawn from our proposed datasets. Topmost - A1 dataset, middle - A2, bottom - A3. }
    \label{fig:in-distribution-results}
\end{figure}

We evaluate \metagen's performance using Relative Error, reported for each of A1, A2, and A3 datasets. The results are averaged over $N=500$ data points which the model has not observed during training. The results are detailed in Table \ref{tab:in-distribution-results}, demonstrating the superiority of \metagen\ in capturing the data distribution. Representative visualized samples are shown in Figure \ref{fig:in-distribution-results}. 

\begin{table}[h]
    \centering
    \begin{tabular}{c|c|c|c}
        \hline
         Model / Dataset      &  A1 & A2 & A3   \\%& NRMS Error $\downarrow$ (Eq. \ref{eq:nrms})\\
         \hline
         CVAE      & $0.5326 \pm 0.2718$ & $0.5951 \pm 0.2017$ & $0.6266 \pm 0.1699$ \\
         WGAN-GP    & $0.2857 \pm 0.1679$ & $0.4655 \pm 0.1725$ & $0.5323 \pm 0.1535$ \\
         \metagen\ (Ours)   & \textbf{0.1012 $\pm$ 0.0926} & \textbf{0.1658 $\pm$ 0.0883} & \textbf{0.2725 $\pm$ 0.1255} \\% & 
         \hline
    \end{tabular}
    \vspace{1em}
    \caption{\small\textbf{Evaluation on the Test Set}. Relative Error comparisons across the test set for various methods, under the physical settings A1, A2, and A3. The results are averaged across 500 samples. Sampling was done without RCWA guidance.}
    \label{tab:in-distribution-results}
\end{table}

\subsection{Generalization to Target Patterns} \label{sec:generalization-to-target}

\begin{figure}[h!]
    \centering
    \begin{subfigure}{1\textwidth}
        \centering
        \includegraphics[width=\textwidth]{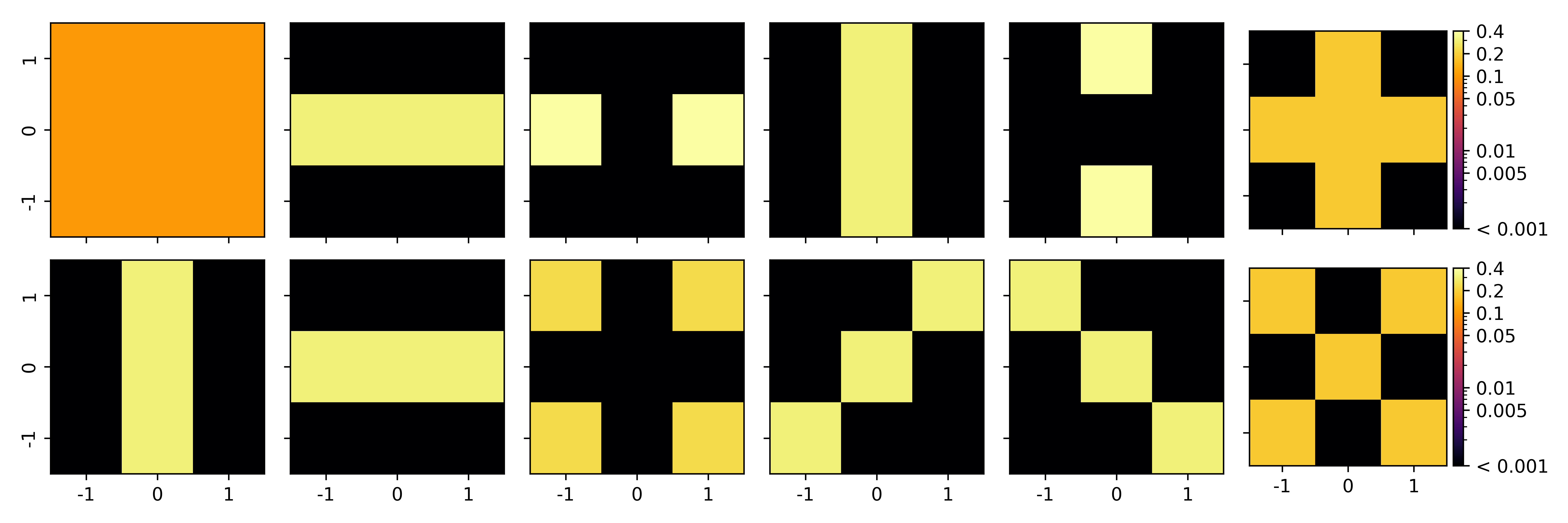}
    \end{subfigure}
    \begin{subfigure}{1\textwidth}
         \centering
         \includegraphics[width=\textwidth]{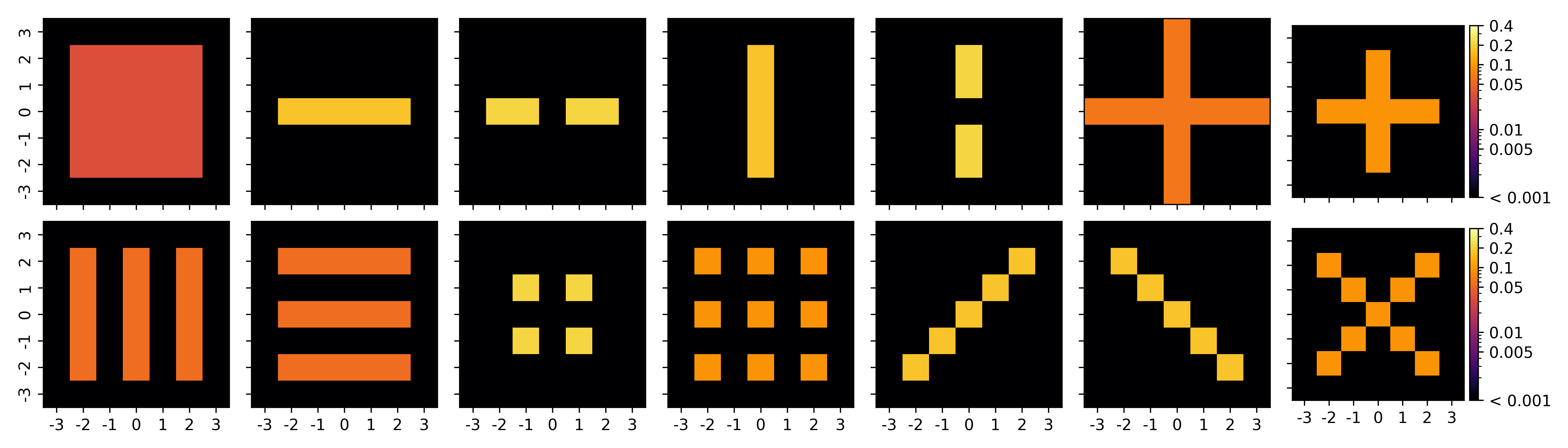}
    \end{subfigure}
    \begin{subfigure}{1\textwidth}
         \centering
         \includegraphics[width=\textwidth]{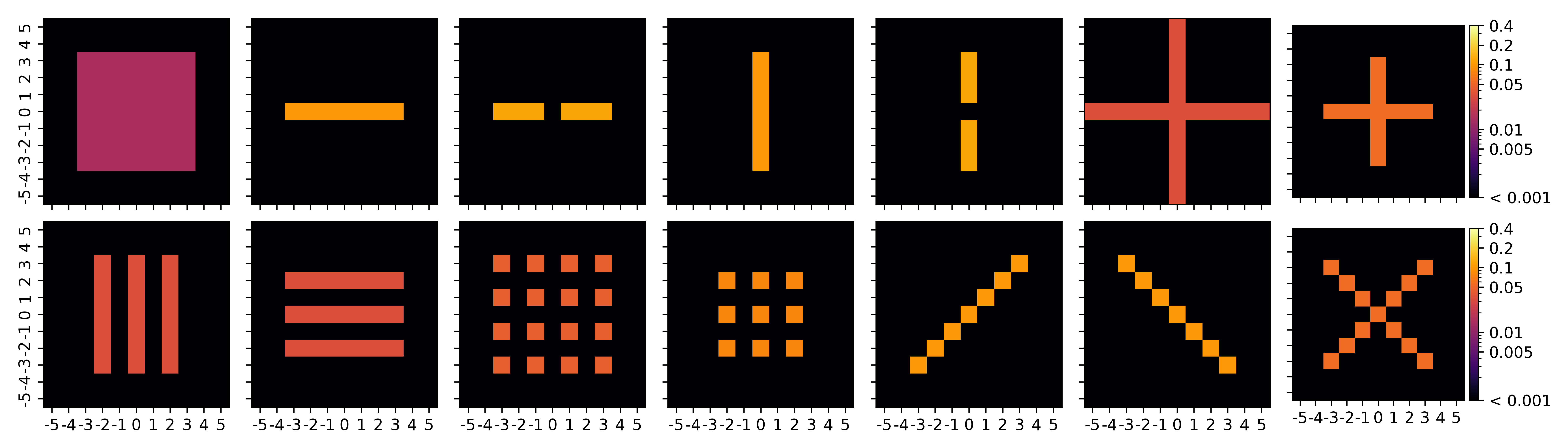}
     \end{subfigure}
    \caption{\small\textbf{Target Patterns}. Top - A1, Middle -A2, Bottom - A3. Each dataset is characterized by a distinct periodicity, hence a different the number of diffraction orders. Accordingly, target patterns were tailored for each dataset.}
    \label{fig:target-pattern-a123}
\end{figure}

Our approach extends beyond training a diffusion model to generate high-fidelity samples as it also synthesizes the data distribution itself. This can result in general but non-functional samples. To address this, we evaluate the model's ability to generate metasurfaces for out-of-distribution patterns, i.e. diffraction patterns that are not included in the training/test sets and might not be fully realizable. For each dataset A1, A2 and A3, we manually construct a reference set of target patterns on which we evaluate all approaches. The target sets are presented in Figure \ref{fig:target-pattern-a123}. It should be emphasized that obtaining precisely these diffraction patterns could prove to be impossible within the constraints (binary, phase-only, metasurfaces with a minimal permittable feature size). Thus, the evaluation of our diffusion model in such cases would be its ability to generate designs that approximate the desired diffraction patterns.

\begin{figure}[h!]
    \centering
    \includegraphics[width=1\linewidth]{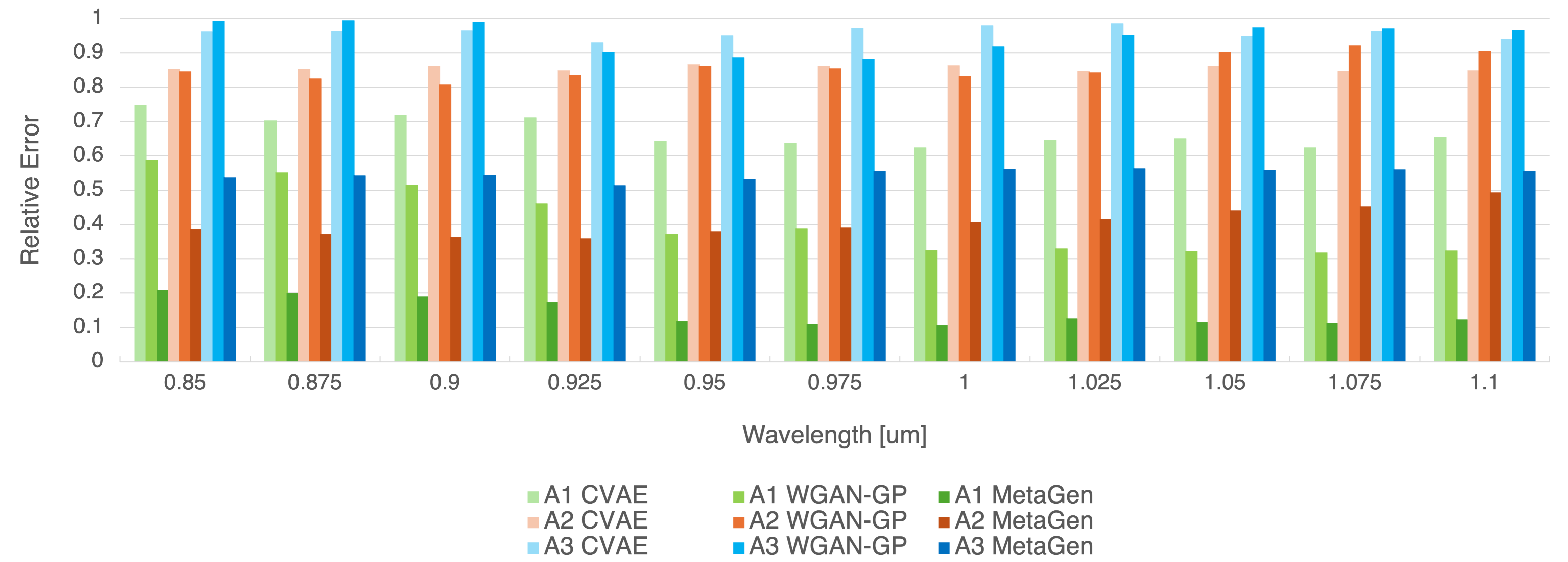}
    \caption{\small \textbf{Evaluation on Target Patterns}. Performance comparison of all examined methods on manually assembled scattering conditions, which lie outside the training data distribution, for each dataset A1, A2, and A3. These conditions are evaluated across multiple operating wavelengths, both included in the training dataset and beyond. The reported results are averaged across 10 experiments for each model and dataset. Sampling is done without RCWA guidance.}
    \label{fig:ood-results}
\end{figure}

\begin{figure}[h!]
    \centering
    \includegraphics[width=\textwidth]{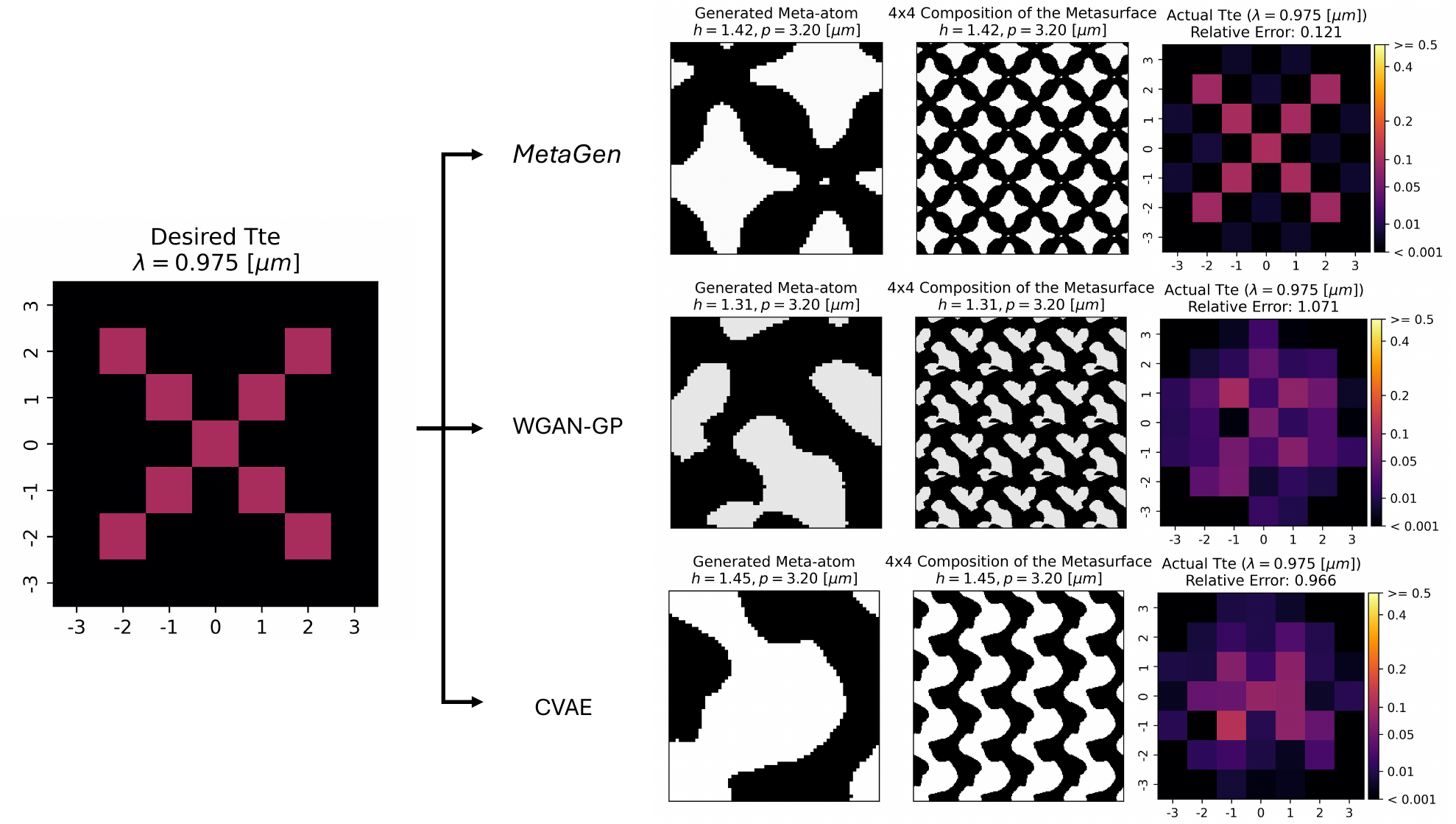}
    \caption{\small\textbf{Target Pattern Sampling}. (Top to bottom) \metagen, WGAN-GP and CVAE sampled metasurfaces conditioned by the presented X-shaped scattering pattern, with out-of-distribution wavelength $\lambda = 0.975 [\mu m]$. The metasurfaces were obtained by sampling each model 10 times and taking the metasurfaces with the lowest relative error. The leftmost and rightmost panels depict, respectively, the desired and obtained spatial power distributions.}
    \label{fig:x-diffraction-a2}
\end{figure}

\hfill \break
We also assess the ability of our approach to generalize by evaluating each method's performance on the target patterns across a broad wavelength range, including wavelengths beyond the training data. Quantitative results are presented in Figure ~\ref{fig:ood-results}. Although each method maintains, roughly, similar accuracy across operating wavelengths within and beyond the training datasets, Figures \ref{fig:ood-results} and \ref{fig:x-diffraction-a2} show that WGAN-GP and CVAE models struggle to generalize from the training data distribution to target patterns. In contrast, \metagen\ not only significantly outperforms the competing methods but also demonstrates strong generalization to out-of-distribution scattering patterns. For example, Figure \ref{fig:x-diffraction-a2} illustrates the results of sampling \metagen\, WGAN-GP, and CVAE 10 times, by feeding the models with both an out-of-distribution scattering pattern and  $\lambda=0.975 [\mu m]$. From this set, the metasurface with the minimal relative error is selected, directly demonstrating the superiority of \metagen\ over competing generative methods. 
%Additional visualizations of metasurfaces generated by \metagen\ in response to target patterns conditioning are presented in Figure \ref{fig:additional-metagen-ood} and in the supplementary material, Figure S5.

\section{Implementation Details}\label{app:implementation-details}

\subsection{\metagen's Implementations Details} \label{app:metagen-details}

\subsubsection{Training}
\begin{figure}[h]
    \centering
    \includegraphics[width=1\linewidth]{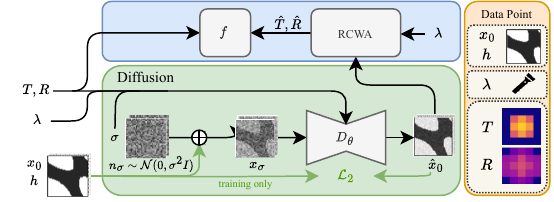}
    \caption{\small\textbf{\metagen\ Training}. Images from the source distribution $x_0$ are noised by a Gaussian noise with zero-mean and random noise level $\sigma$, then passed to the diffusion model which is conditioned by both the noise level and the scattering patterns.}
    \label{fig:training}
\end{figure}

We train a diffusion model, implemented using EDM, the pre-conditioning and sampling framework proposed in \cite{EDM}. The training process is illustrated in Figure~\ref{fig:training} and is explained hereby. The model receives $x_0$, a sample from the source distribution, augmented by random cyclic-shifts, along with its corresponding height $h$, both concatenated together channel-wise to form a $[2, 64, 64]$ shaped Tensor. It is then noised by a Gaussian noise $n_\sigma \sim \mathcal{N}(0, \sigma^2I)$ with $\sigma$ drawn from a log-normal distribution, i.e. $\log\sigma \sim \mathcal{N}(\mu, s^2)$ for some proper choices of $\mu, s$ (see the supplementary material, Section 2.A). The noised sample, denoted by $x_\sigma := x_0 + n_\sigma$, is fed into the diffusion model, which is a U-Net shaped encoder-decoder neural network \cite{unet, EDM}. The scattering patterns and the wavelength $\lambda$, along with the noise level $\sigma$, are fed into the diffusion model as conditional signals and later injected into each level of the encoder and decoder. The model is trained to reconstruct a denoised version of $x_\sigma$, denoted as $\hat{x}_0$, conditioned by the additional input parameters. We apply the $\mathcal{L}_2$ loss criterion between the target image $x_0$ and its reconstruction $\hat{x}_0$. Because we focus on the transmitted spatial power distribution $T$, we use $T$ and $\lambda$ only as conditional signals, and omit $R$. 

\paragraph{Noise level distribution.} 
Contrary to the common practice of sampling $\sigma$ uniformly, empirical experiments done in \cite{EDM} suggest sampling $\sigma$ from a log-normal distribution, i.e $\log\sigma \sim \mathcal{N}(\mu, s^2)$, with $\mu, s$, optimally found for natural images to be $-1.2, 1.2$ respectively, implying $\sigma$ values between $0.09$ to $1$ within one standard deviation. We follow this spirit, but rather set $\mu = 0$ to induce higher noise levels ranging from $0.3$ to $3.3$ within one standard deviation. This adjustment is based on the observation that denoising binary images, as well as constant images, is less challenging than denoising natural images.

\paragraph{Cyclic augmentations.}Due to the periodic nature of the overall metasurface, the intensity of the transmitted and reflected light is invariant to cyclic shifts of the meta-atom topology. This fact facilitates data augmentations, which expand the dataset and ultimately enhance the model’s generative capabilities. Moreover, as the model is trained by observing meta-atoms only, we hypothesize that applying cyclic-shifts augmentations informs the model of the periodic nature of these kinds of meta-surfaces.

In practice, a random cyclic shift is applied to the input binary image during training while keeping the corresponding scattering pattern unchanged. We note that during training, the model can partially reconstruct moderately noisy samples. However, when presented with an extremely noisy sample, $x_\sigma$, which contains almost no information, the model might ignore the noisy input and rely solely on the conditional information. This can result in the generation of a new meta-atom, $\hat{x}_0$, which may be a cyclic-shifted variant of $x_0$, the original, clean binary structure. In such cases, due to the cyclic pixel-wise misalignment, the model may incur a high $\mathcal{L}_2$ loss penalty, causing it to deviate from a desired behavior. To mitigate this, we apply cyclic-shift correction by aligning the denoised structure $\hat{x}_0$ to the original clean structure $x_0$. This correction is performed by computing the maximal correlation between $x_0$ and $\hat{x}_0$, implemented using a 2-D Fast Fourier Transform for efficiency, and cyclically shifting the image accordingly.

\paragraph{Training parameters.} All \metagen\ models have  ~$62M$ parameters, trained with batch size of 32 and optimized by AdamW Schedule-Free optimizer \cite{adamw, schedulefree} with learning rate $10^{-4}$ and no weight decay regularization. The models for datasets B2 and C2 were trained for $3M$ optimization steps, while the models for A1, A2 and A3 were trained for $1M$ steps.

\subsubsection{Sampling Details}\label{app:sampling-details}

The authors of EDM~\cite{EDM} proposed a noise schedule over discretized time steps that achieves strong sampling performance on natural image datasets. The schedule defines a decreasing sequence of noise levels for \( i \in \{0, \dots, T\} \), where earlier time steps correspond to higher noise magnitudes:

\begin{equation}\label{eq:edm_noise_schedule}
    \sigma_i = \begin{cases}
        \left(\sigma_{\text{max}}^{1/\rho} + \dfrac{i}{T-1} \left(\sigma_{\text{min}}^{1/\rho} - \sigma_{\text{max}}^{1/\rho}\right)\right)^{\rho}, & \text{if } i < T \\
        0, & \text{if } i = T
    \end{cases}
\end{equation}

While this schedule is effective for natural image generation, our setting involves a joint distribution over binary metasurface structures and their corresponding scattering patterns, which differs significantly in structure and semantics. Accordingly, although we retain the overall shape of the noise schedule, we adapt its parameters to better suit our data. In particular, we slow down the progression at high noise levels, where conditional information (e.g., the desired scattering pattern) is parsed, and accelerate it at low noise levels, where the model primarily refines fine structural details.

The original EDM parameters are provided in~\cite{EDM}. In our implementation, we increase \( \sigma_{\text{max}} \) to 100 and set \( \rho = 3 \), encouraging more gradual transitions in the early (noisy) stages, which was found to improve performance empirically. We use \( T = 100 \) time steps and follow the sampling procedure described in Algorithm 2 of~\cite{EDM}, with the following key modifications:

\begin{itemize}
    \item \textbf{Deterministic Sampling.} We disable stochasticity by setting \( S_{\text{churn}} = 0 \), ensuring a stable and consistent sampling trajectory given the initial noise $n_\sigma$.
    
    \item \textbf{Posterior Sampling via DPS (RCWA Guidance).} We incorporate Diffusion Posterior Sampling (DPS)~\cite{chung2022diffusion}, which introduces a measurement consistency term into the sampling process. Specifically, we modify the intermediate noised sample at each step by adding a gradient correction term that minimizes the \( \mathcal{L}_2 \) error between the predicted and desired scattering patterns. We find that scaling this gradient by the noise level \( \sigma_i \) at each timestep $i$ yields the best empirical results.
    
    \item \textbf{First-Order Solver.} We omit the second-order Heun correction term proposed in~\cite{EDM}, as we observe no performance gain in our setting. Furthermore, we find that this first-order Euler method integrates more naturally with DPS-guided sampling.
\end{itemize}

\subsubsection{Architecture}
\metagen\ adopts the U-Net backbone architecture introduced in~\cite{SongDiffusion}, as implemented in EDM~\cite{EDM}. The architecture of \metagen\ features a mirror-like encoder-decoder structure composed of UNet blocks, detailed in Table \ref{unet-block}. Skip connections link the outputs of each encoder block to their corresponding decoder blocks. Conditional information is incorporated by embedding an $N$-dimensional vector into a high-dimensional space and injecting it into each UNet block throughout the network via element-wise addition at every level. The complete architecture of \metagen\, including the shapes of all intermediate components, is outlined in Table \ref{neural_network}.

\begin{table}[h!]
\small
\centering
\begin{tabular}{|c|l|c|c|l|}
\hline
\textbf{Layer \#}& \textbf{Type}      & \textbf{Inputs (by Layer \#)}& \textbf{Output Shape}          &\textbf{Notes}\\ \hline
1                 & Inputs& $x$& $[B, C_\text{in}, H, W]$&\\
2 & & $emb$& $[B, K]$&\\\hline
3& Conv2D& 1& $[B, C_\text{out}, H, W]$&SiLU Activation\\ \hline
 4& Linear& 2& $[B, C_\text{out}]$&\\ \hline
5& Group Norm& 3, 4& $[B, C_\text{out}, H, W]$      &SiLU Activation\\ \hline
6& Conv2D& 5& $[B, C_\text{out}, H, W]$&\\ \hline
7& Group Norm& 6, 1& $[B, C_\text{out}, H, W]$      &\\ \hline
8& Self-Attention& 7& $[B, C_\text{out}, H, W]$      &\\ \hline
9& Conv2D& 8& $[B, C_\text{out}, H, W]$      &\\ \hline
10& Output             & 9, 6& $[B, C_\text{out}, H, W]$      &\\ \hline
\end{tabular}
\vspace{1em}
\caption{\small\textbf{UNet Block($C_\text{in}, C_\text{out}, K$) Description}. The block has two inputs: an image $x$ of shape $[H,W]$ where $C_\text{in}$ is the number of channels, and a $K$-dimensional embedding $emb$ that encodes the conditional information. Skip connections between sequential sub-modules within the block are implemented as element-wise additions.}
\label{unet-block}
\end{table}

\pagebreak

\begin{table}[h]
\small
\centering
\begin{tabular}{|c|l|>{\centering\arraybackslash}p{1.5cm}|c|l|}
\hline
\textbf{Layer \#} & \textbf{Type}             & \textbf{Inputs (by Layer \#)} & \textbf{Output Shape}         & \textbf{Notes}\\ \hline
 1& Inputs& $x$& $[B, 2, 64, 16]$&\\
 2& & $y$& $[B, N]$&\\
 3& & $\sigma$& $[B,1]$&\\\hline
4& Positional Embedding& 3& $[B, 128]$& -    \\ \hline
5& Linear& 2, 4& $[B, 128]$& -    \\ \hline
6& Linear& 5& $[B, 512]$& SiLU Activation\\ \hline
7& Linear& 6& $[B, 512]$& SiLU Activation\\ \hline
 8& Conv2D& 1& $[B, 2, 64, 64]$&-    \\ \hline
9& 4 x UNetBlock& 8, 7& $[B, 128, 64, 64]$& -\\ \hline
10& Down Sampling UNet Block& 9, 7& $[B, 128, 32, 32]$& -\\ \hline
11& 4 x UNetBlock& 10, 7& $[B, 256, 32, 32]$& -\\ \hline
12& Down Sampling UNet Block& 11, 7& $[B, 256, 16, 16]$& -\\ \hline
13& 4 x UNetBlock& 12, 7& $[B, 256, 16, 16]$& Inc. Self-Attention.\\ \hline
14& Down Sampling UNet Block& 13, 7& $[B, 256, 8, 8]$& -\\ \hline
15& 4 x UNetBlock& 14, 7& $[B, 256, 8, 8]$& -\\ \hline
16& 2 x UNetBlock& 15, 7& $[B, 256, 8, 8]$& Inc.Self-Attention.\\ \hline
17& 5 x UNetBlock& 16, 7, 15& $[B, 256, 8, 8]$& -\\ \hline
18& Up Sampling UNet Block& 17, 7, 14& $[B, 256, 16, 16]$& -\\ \hline
19& 5 x UNetBlock& 18, 7, 13& $[B, 256, 16, 16]$& Inc. Self-Attention.\\ \hline
20& Up Sampling UNet Block& 19, 7, 12& $[B, 256, 32, 32]$& -\\ \hline
21& 5 x UNetBlock& 20 ,7, 11& $[B, 256, 32, 32]$& -\\ \hline
22& Up Sampling UNet Block& 21, 7, 10& $[B, 128, 64, 64]$& -\\ \hline
23& 5 x UNetBlock& 22, 7& $[B, 128, 64, 64]$& -\\ \hline
24& Group Norm& 23& $[B, 128, 64, 64]$& SiLU Activation\\ \hline
25& Conv2D& 24& $[B, 2, 64, 16]$& -\\ \hline
\end{tabular}
\vspace{1em}
\caption{\small Our diffusion model is inherited from~\cite{EDM} and follows the DDPM++ architecture proposed in \cite{SongDiffusion}. The network adopts a mirror-like Encoder-Decoder structure. Skip connections are implemented by concatenating the encoder outputs with inputs of its corresponding decoder layer. The self-attention layer in the UNet blocks is omitted by default unless explicitly stated otherwise. Down and up sampling are achieved using stride $=2$ in convolution and transposed convolution layers, respectively.}
\label{neural_network}
\end{table}

\subsection{WGAN-GP Implementation Details} \label{app:gan-details}
To evaluate our model, we trained a Wasserstein GAN with Gradient Penalty (WGAN-GP) \cite{arjovsky2017wasserstein, gulrajani2017improved, mirza2014conditional,Cohen2023GAN} on each of our datasets (A1, A2, and A3). To adapt WGAN-GP for our conditional generation task, we modified the standard framework by conditioning both the Generator and Critic networks on the scattering patterns. Conditioning in the Generator was achieved by incorporating a small 3-layer Multi-Layer Perceptron (MLP) to embed the scattering patterns into a learned high-dimensional space, which was then concatenated with random noise, as is typical in GAN architectures. For the Critic, the same embedding MLP was used, with its output reshaped to match the spatial dimensions of the input image. The embedding dimensionality for both networks was thus set to $64^2$, and the latent noise dimension was set to 100. 
The training process on each of our dataset A1, A2 and A3, consisted of 1M optimization steps for the Generator, and for each Generator update, a decoupled separated optimization step was performed for the Critic. The batch size was set to 64. Both the Generator and the Critic were optimized by Adam optimizer \cite{adam} with unscheduled learning rate of $0.0001$, and $\beta_1 = 0, \beta_2 = 0.9$.
These training settings were selected to align closely with the training procedure of \metagen\ for a fair comparison. Additionally, the embedding MLP used for conditioning both the Generator and the Critic was identical to the one used in \metagen, following the implementation of \cite{EDM}.

\subsection{C-VAE Implementation Details} \label{app:vae-details}
Building upon an existing implementation of a Conditional Variational Autoencoder (C-VAE) \cite{sohn2015learning,Bank2023AE}, we introduced architectural enhancements to ensure competitiveness with our method. Standard implementations of Conditional VAEs typically condition the model by providing the auxilary information as input only to the first layer of the Encoder and Decoder. For enhancing the compliance of the model to the conditional information, we rather utilize a preceding MLP to first embed this information into a learned high-dimensional space, and inject it into every convolutional layer of both Encoder and Decoder networks by stacking the embedded conditional information as an extra data channel. Moreover, the embedded conditional information is incorporated into the final layer of the Encoder, which outputs the mean and variance of the learned posterior distribution of the latent space. This design enforces a strong impact of the conditions on the structure of the latent space.
To complement these enhancements, we added several unconditional residual layers in cascade to both the Encoder and the Decoder, creating of a more meaningful latent space. Together, these improvements significantly strengthen the enforcement of conditional information and the model’s overall performance.

We trained a Conditional Variational Autoencoder (C-VAE) separately for each dataset - A1, A2, and A3 - over 200 epochs with a batch size of 1024. The training employed Adam optimizer with a learning rate of $0.001$, a decay rate of $0.95$, and no weights decay. The Kullback-Leibler Divergence loss term was weighted by a factor of $0.00025$ to balance its impact on the overall loss function.The latent space dimension was set to 100.

\section{Gradient Descent for Metasurface Inverse Design} \label{app:gd-details}
For metasurface inverse design, standard gradient descent (GD) algorithms must be adapted to produce binary outputs while maintaining fabrication feasibility, i.e. eliminating small features and preserving periodic continuity. In this study, we follow the approach introduced in~\cite{Kim2020diffraction, torcwa}, modifying the GD scheme to suit metasurface design by incorporating two key components: a fixed spatial blurring kernel and a soft, differentiable binarization function that is gradually sharpened during optimization.

The blurring step is applied in the spatial frequency domain using the filter defined in Eq.~\ref{eq:lpf}, as described in Section~\ref{app:data} of this supplementary material. To enforce binarization while maintaining gradient flow, we apply a soft projection function with a time-dependent sharpness parameter. This function is defined as:

\begin{equation}\label{eq:gradual-thresholding}
    \mathcal{P}(x; t) = \frac{\tanh\left(\frac{1}{2}\beta(t)\right) + \tanh\left(\left(x - \frac{1}{2}\right)\beta(t)\right)}{2\tanh\left(\frac{1}{2}\beta(t)\right)}, 
    \quad \text{where} \quad \beta(t) = \exp\left(\frac{t}{N} \ln(\beta_{\text{max}})\right)
\end{equation}

where \( N \) is the total number of optimization steps, and \( \beta_{\text{max}} = 1000 \) is the final binarization sharpness. The projection function \( \mathcal{P}(x; t) \) smoothly approximates a step function, becoming increasingly sharper as \( t \) increases, thereby gradually enforcing binary values while preserving differentiability throughout the optimization.

\section{Hardware Statement} \label{app:hardware}
All computations in this study were performed using a single NVIDIA L40S GPU, utilizing the GPU-accelerated PyTorch framework. The codebase relies on GPU support not only for model training, but also for sampling, optimization, and forward measurements conducted via the ToRCWA simulator~\cite{torcwa}. These components are implemented to leverage the computational efficiency of modern GPUs. While the code is compatible with CPU execution, doing so requires sufficient memory and results in significantly slower runtimes.

\section{Additional Results}\label{app:additional-results}
This section presents additional examples of metasurfaces generated by \metagen\ for qualitative evaluation. Figure~\ref{fig:ood-additional-results} shows results conditioned on target patterns constructed for datasets A1, A2, and A3. These examples highlight \metagen’s ability to generate metasurface designs that match diffraction patterns beyond the range of the training set, both in terms of spatial configuration and wavelength.

\begin{figure}[h]
    \centering
     \begin{subfigure}{1\textwidth}
         \centering
         \includegraphics[width=\textwidth]{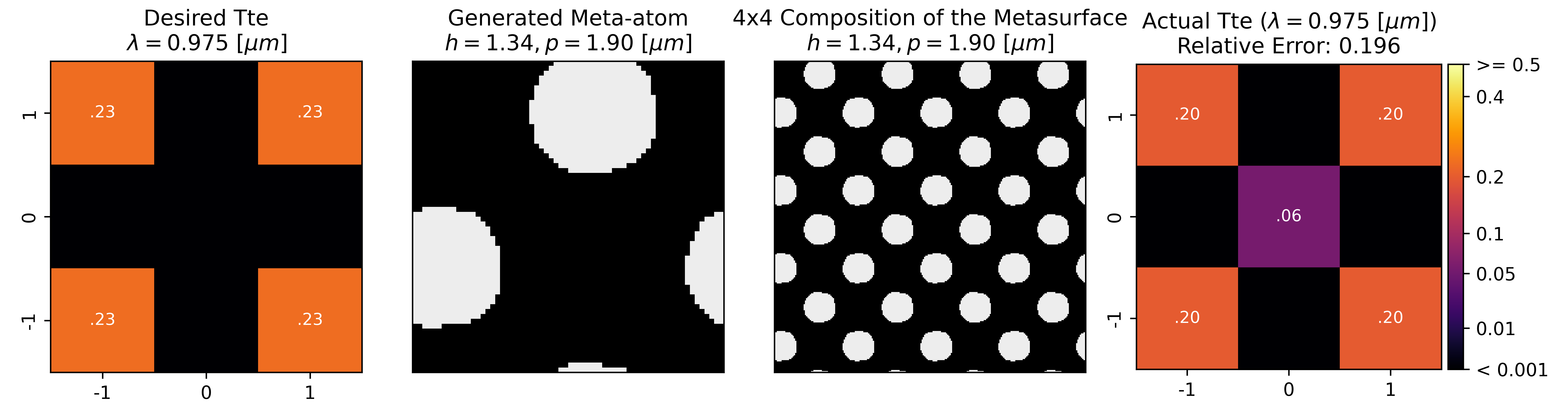}
     \end{subfigure}
\end{figure}
\begin{figure}\ContinuedFloat
    \begin{subfigure}{1\textwidth}
         \centering
         \includegraphics[width=\textwidth]{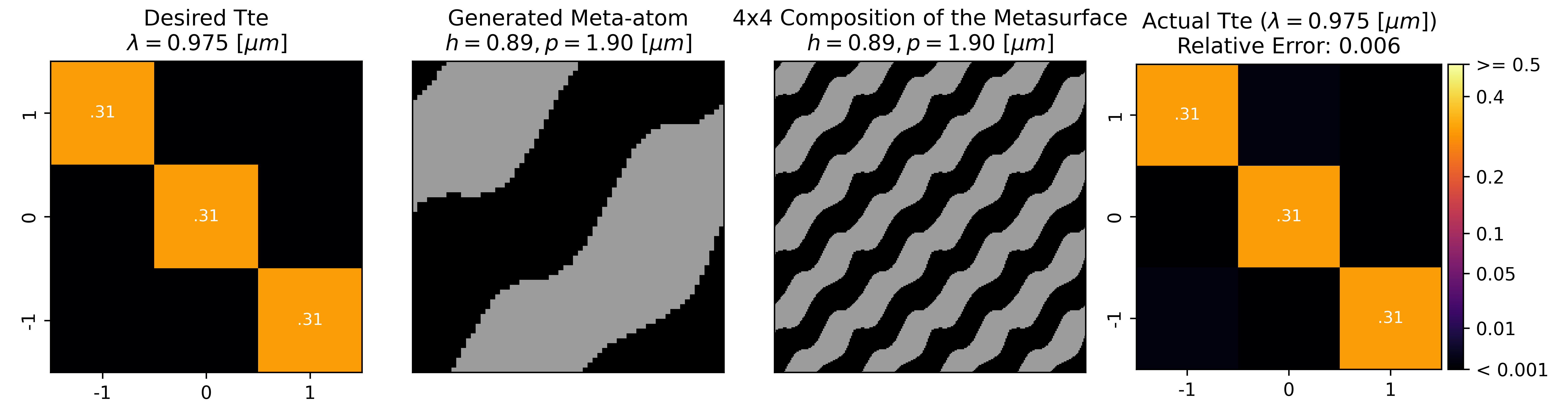}
     \end{subfigure}
     \begin{subfigure}{1\textwidth}
         \centering
         \includegraphics[width=\textwidth]{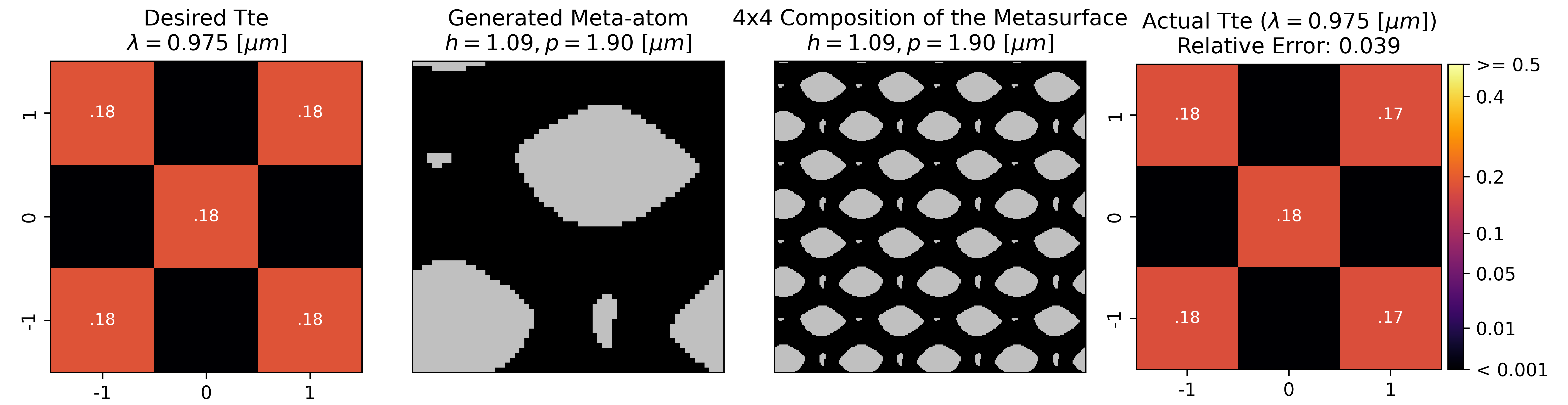}
     \end{subfigure}
     \begin{subfigure}{1\textwidth}
         \centering
         \includegraphics[width=\textwidth]{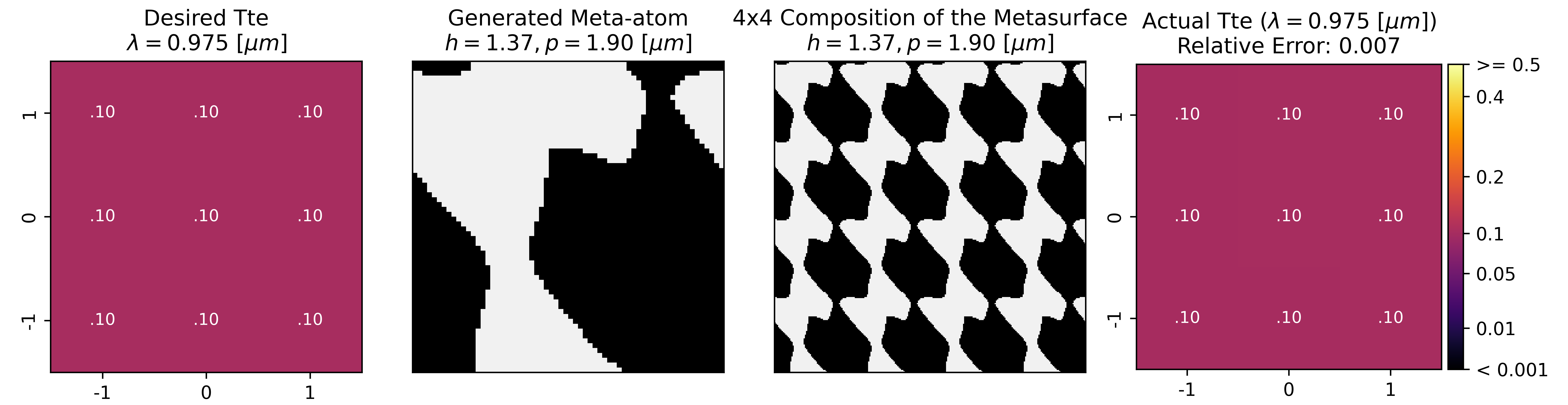}
     \end{subfigure}
     \begin{subfigure}{1\textwidth}
         \centering
         \includegraphics[width=\textwidth]{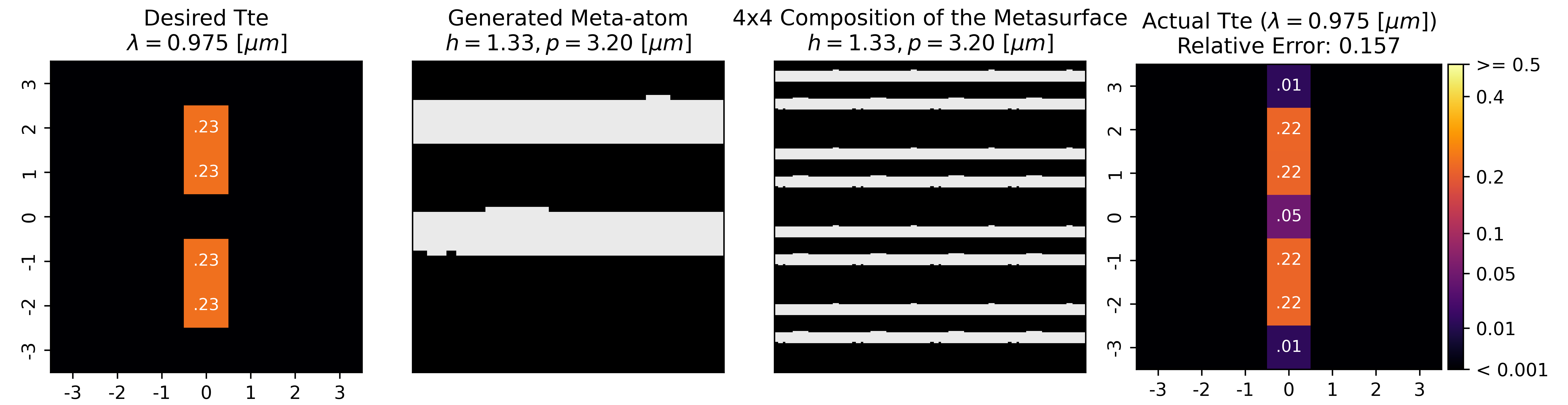}
     \end{subfigure}
     \begin{subfigure}{1\textwidth}
         \centering
         \includegraphics[width=\textwidth]{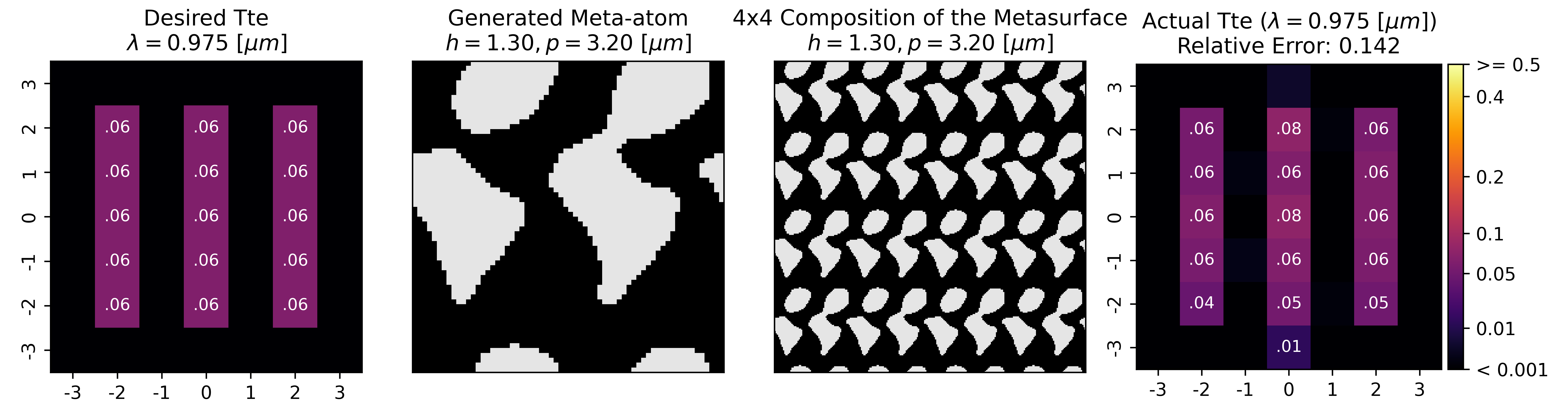}
     \end{subfigure}
\end{figure}
\begin{figure}\ContinuedFloat
     \begin{subfigure}{1\textwidth}
         \centering
         \includegraphics[width=\textwidth]{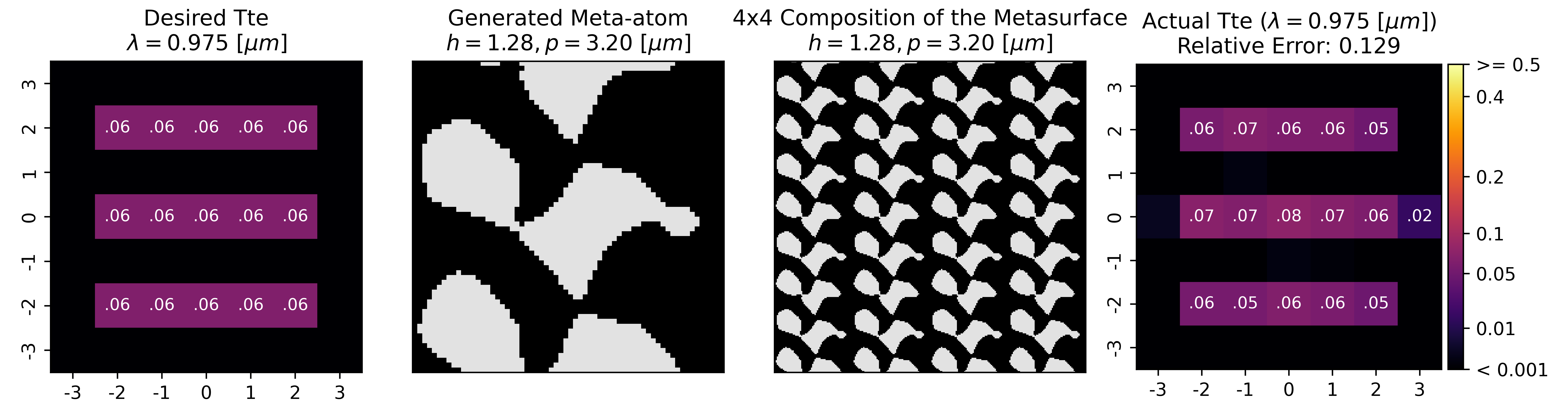}
     \end{subfigure}
     \begin{subfigure}{1\textwidth}
         \centering
         \includegraphics[width=\textwidth]{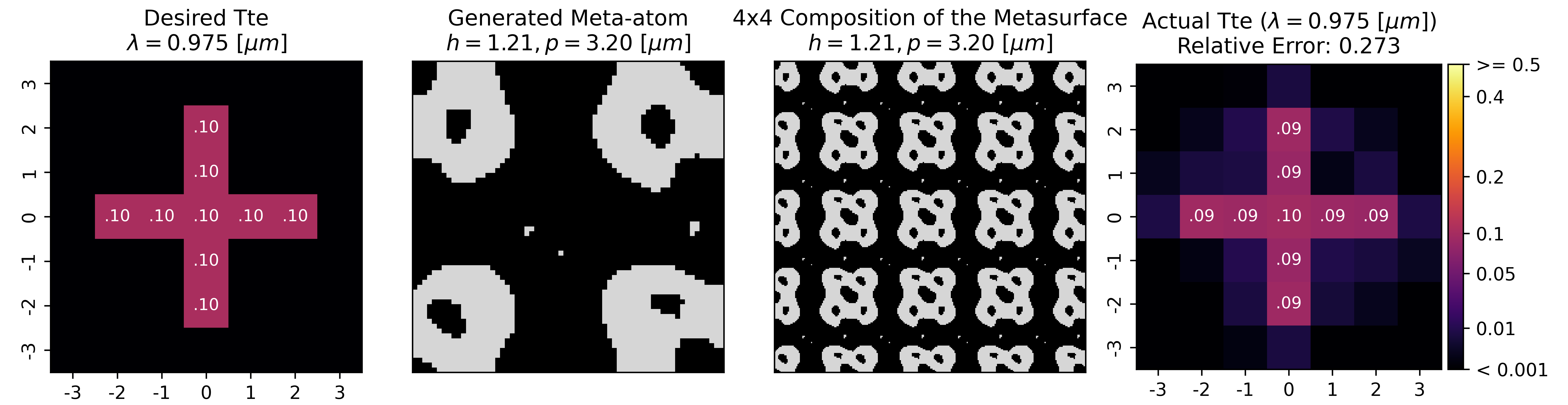}
     \end{subfigure}
     \begin{subfigure}{1\textwidth}
         \centering
         \includegraphics[width=\textwidth]{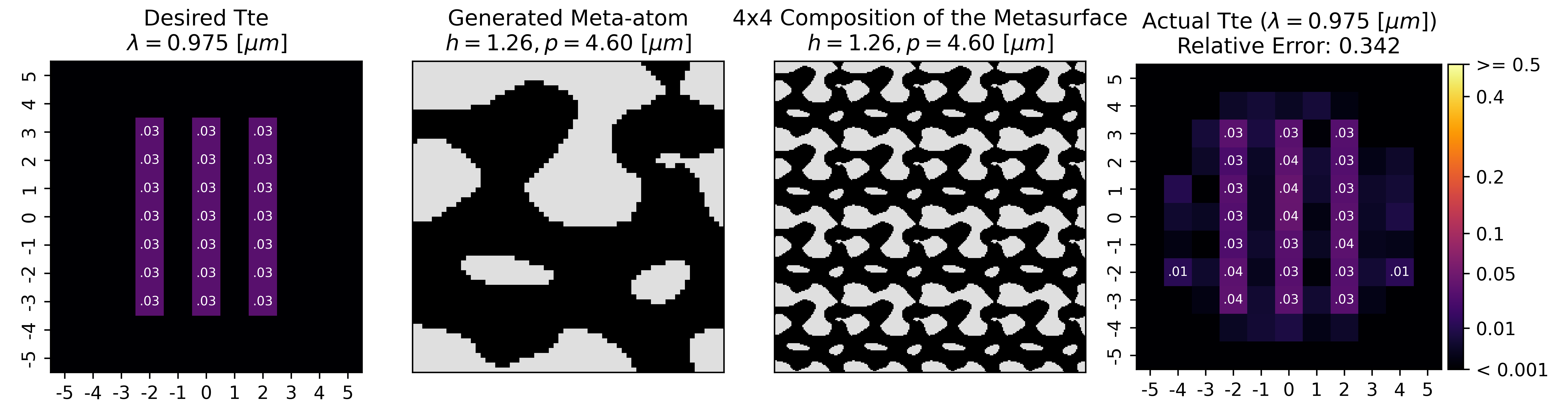}
     \end{subfigure}
     \begin{subfigure}{1\textwidth}
         \centering
         \includegraphics[width=\textwidth]{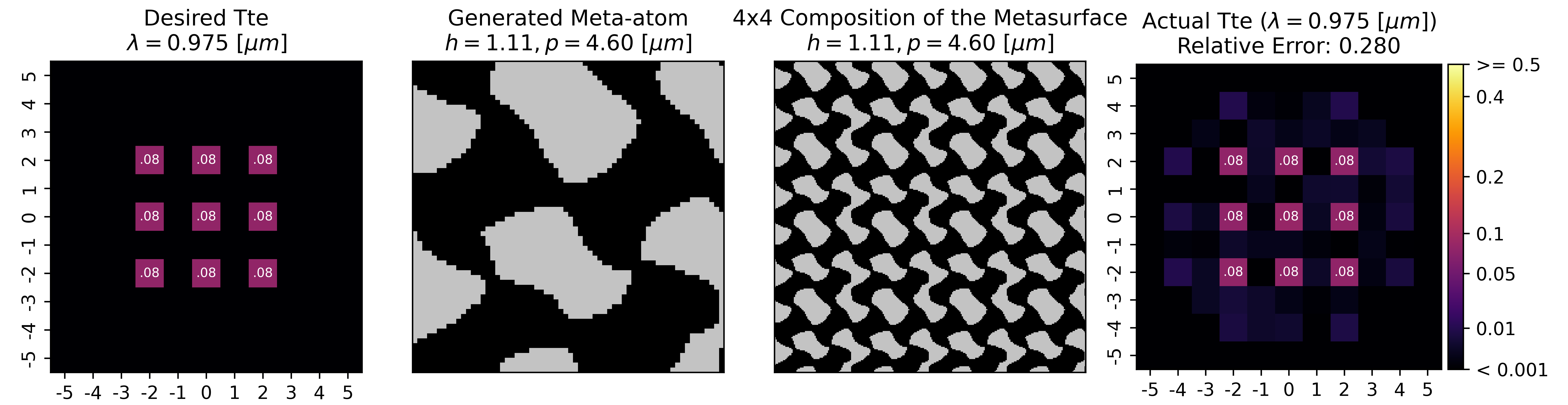}
     \end{subfigure}
     \begin{subfigure}{1\textwidth}
         \centering
         \includegraphics[width=\textwidth]{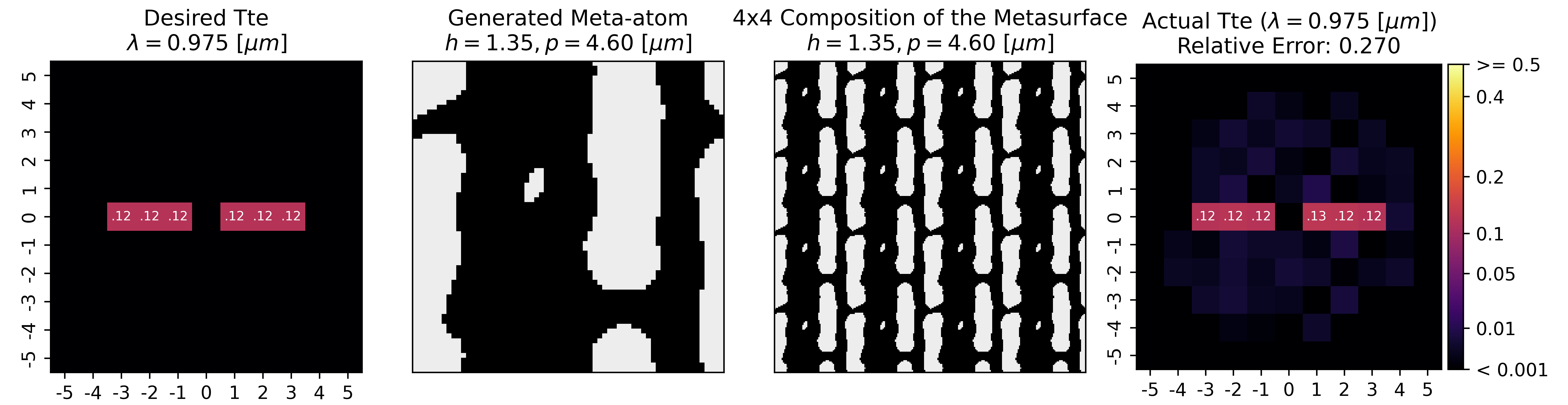}
     \end{subfigure}
    \caption{\small\textbf{Representative Results For Out-of-distribution Samples.} The examples are generated from all \metagen\ models (A1, A2 and A3), all under illumination of $\lambda = 0.975 [\mu m]$ which was not included in the training data.}
    \label{fig:ood-additional-results}
\end{figure}


\begin{thebibliography}{10}

\bibitem{RCWAMoharam95}
M.~G. Moharam, Eric~B. Grann, Drew~A. Pommet, and T.~K. Gaylord.
\newblock Formulation for stable and efficient implementation of the rigorous coupled-wave analysis of binary gratings.
\newblock {\em J. Opt. Soc. Am. A}, 12(5):1068--1076, May 1995.

\bibitem{torcwa}
Changhyun Kim and Byoungho Lee.
\newblock Torcwa: Gpu-accelerated fourier modal method and gradient-based optimization for metasurface design.
\newblock {\em Computer Physics Communications}, 282:108552, 2023.

\bibitem{Peurifoy2018}
John Peurifoy, Yichen Shen, Li~Jing, Yi~Yang, Fidel Cano-Renteria, Brendan~G. DeLacy, John~D. Joannopoulos, Max Tegmark, and Marin Soljačić.
\newblock Nanophotonic particle simulation and inverse design using artificial neural networks.
\newblock {\em Science Advances}, 4(6):eaar4206, 2018.

\bibitem{Kim2020diffraction}
Dong~Cheon Kim, Andreas Hermerschmidt, Pavel Dyachenko, and Toralf Scharf.
\newblock Inverse design and demonstration of high-performance wide-angle diffractive optical elements.
\newblock {\em Opt. Express}, 28(15):22321--22333, Jul 2020.

\bibitem{TopologyOptimization4InverseDesign}
Rasmus~E. Christiansen and Ole Sigmund.
\newblock Inverse design in photonics by topology optimization: tutorial.
\newblock {\em J. Opt. Soc. Am. B}, 38(2):496--509, Feb 2021.

\bibitem{EvolutionaryAlgorithms4InverseDesign}
Zhaoyi Li, Raphaël Pestourie, Zin Lin, Steven~G. Johnson, and Federico Capasso.
\newblock Empowering metasurfaces with inverse design: Principles and applications.
\newblock {\em ACS Photonics}, 9(7):2178--2192, 2022.

\bibitem{hao2019diffraction}
Huang Hao, Zhai Tingting, Song Qiang, and Yin Xiaodong.
\newblock Wide angle 2d beam splitter design based on vector diffraction theory.
\newblock {\em Optics Communications}, 434:28--35, 2019.

\bibitem{Li2021diffraction}
Jinzhe Li, Fei Zhang, Mingbo Pu, Yinghui Guo, Xiong Li, Xiaoliang Ma, Changtao Wang, and Xiangang Luo.
\newblock Quasi-continuous metasurface beam splitters enabled by vector iterative fourier transform algorithm.
\newblock {\em Materials}, 14(4), 2021.

\bibitem{Yan2023Diffraction}
Yu~Yan, Zhentao Fan, Guofang Sun, and Kehan Tian.
\newblock {Diffractive optical element design based on vector diffraction theory and improved PSO-SA algorithm}.
\newblock {\em Optical Engineering}, 62(2):025103, 2023.

\bibitem{tanriover2023generative}
Ibrahim Tanriover, Doksoo Lee, Wei Chen, and Koray Aydin.
\newblock Deep generative modeling and inverse design of manufacturable free-form dielectric metasurfaces.
\newblock {\em ACS Photonics}, 10(4):875--883, 2023.

\bibitem{Diffusion4InverseDesign1}
Zezhou Zhang, Chuanchuan Yang, Yifeng Qin, Hao Feng, Jiqiang Feng, and Hongbin Li.
\newblock Diffusion probabilistic model based accurate and high-degree-of-freedom metasurface inverse design.
\newblock {\em Nanophotonics}, 12(20):3871–3881, October 2023.

\bibitem{Diffusion4InverseDesign2}
Chen Niu, Mario Phaneuf, and Puyan Mojabi.
\newblock A diffusion model for multi-layered metasurface unit cell synthesis.
\newblock {\em IEEE Open Journal of Antennas and Propagation}, 4:654--666, 2023.

\bibitem{Diffusion4InverseDesign3}
Zezhou Zhang, Chuanchuan Yang, Yifeng Qin, Zhihai Zheng, Jiqiang Feng, and Hongbin Li.
\newblock Addressing high-performance data sparsity in metasurface inverse design using multi-objective optimization and diffusion probabilistic models.
\newblock {\em Opt. Express}, 32(23):40869--40885, Nov 2024.

\bibitem{SongDiffusion}
Yang Song, Jascha Sohl-Dickstein, Diederik~P. Kingma, Abhishek Kumar, Stefano Ermon, and Ben Poole.
\newblock Score-based generative modeling through stochastic differential equations, 2021.

\bibitem{ddim}
Jiaming Song, Chenlin Meng, and Stefano Ermon.
\newblock Denoising diffusion implicit models.
\newblock {\em arXiv preprint arXiv:2010.02502}, 2020.

\bibitem{EDM}
Tero Karras, Miika Aittala, Timo Aila, and Samuli Laine.
\newblock Elucidating the design space of diffusion-based generative models.
\newblock In {\em Proc. NeurIPS}, 2022.

\bibitem{chung2022diffusion}
Hyungjin Chung, Jeongsol Kim, Michael~T Mccann, Marc~L Klasky, and Jong~Chul Ye.
\newblock Diffusion posterior sampling for general noisy inverse problems.
\newblock {\em arXiv preprint arXiv:2209.14687}, 2022.

\bibitem{arjovsky2017wasserstein}
Martin Arjovsky, Soumith Chintala, and L{\'e}on Bottou.
\newblock Wasserstein generative adversarial networks.
\newblock In {\em International conference on machine learning}, pages 214--223. PMLR, 2017.

\bibitem{gulrajani2017improved}
Ishaan Gulrajani, Faruk Ahmed, Martin Arjovsky, Vincent Dumoulin, and Aaron~C Courville.
\newblock Improved training of wasserstein gans.
\newblock {\em Advances in neural information processing systems}, 30, 2017.

\bibitem{mirza2014conditional}
Mehdi Mirza.
\newblock Conditional generative adversarial nets.
\newblock {\em arXiv preprint arXiv:1411.1784}, 2014.

\bibitem{Cohen2023GAN}
Gilad Cohen and Raja Giryes.
\newblock {\em Generative Adversarial Networks}, pages 375--400.
\newblock Springer International Publishing, 2023.

\bibitem{sohn2015learning}
Kihyuk Sohn, Honglak Lee, and Xinchen Yan.
\newblock Learning structured output representation using deep conditional generative models.
\newblock {\em Advances in neural information processing systems}, 28, 2015.

\bibitem{Bank2023AE}
Dor Bank, Noam Koenigstein, and Raja Giryes.
\newblock {\em Autoencoders}, pages 353--374.
\newblock Springer International Publishing, 2023.

\bibitem{Dhariwal2021DiffusionMB}
Prafulla Dhariwal and Alex Nichol.
\newblock Diffusion models beat gans on image synthesis.
\newblock {\em ArXiv}, abs/2105.05233, 2021.

\bibitem{DiffusionBeatGANTO}
Fran{\c{c}}ois Maz{\'e} and Faez Ahmed.
\newblock Diffusion models beat gans on topology optimization.
\newblock In {\em Proceedings of the AAAI conference on artificial intelligence}, volume~37, pages 9108--9116, 2023.

\bibitem{anderson1982reverse}
Brian~DO Anderson.
\newblock Reverse-time diffusion equation models.
\newblock {\em Stochastic Processes and their Applications}, 12(3):313--326, 1982.

\bibitem{efron2011tweedie}
Bradley Efron.
\newblock Tweedie’s formula and selection bias.
\newblock {\em Journal of the American Statistical Association}, 106(496):1602--1614, 2011.

\bibitem{hyvarinen2005estimation}
Aapo Hyv{\"a}rinen and Peter Dayan.
\newblock Estimation of non-normalized statistical models by score matching.
\newblock {\em Journal of Machine Learning Research}, 6(4), 2005.

\bibitem{vincent2011connection}
Pascal Vincent.
\newblock A connection between score matching and denoising autoencoders.
\newblock {\em Neural computation}, 23(7):1661--1674, 2011.

\bibitem{ho2022classifier}
Jonathan Ho and Tim Salimans.
\newblock Classifier-free diffusion guidance.
\newblock {\em arXiv preprint arXiv:2207.12598}, 2022.

\bibitem{Loewen2018DiffractionGA}
Erwin~G. Loewen and Evgeny Popov.
\newblock Diffraction gratings and applications.
\newblock In {\em Diffraction Gratings and Applications}, 2018.

\bibitem{pseudo-freeform-dataset}
Sensong An, Clayton Fowler, Bowen Zheng, Mikhail~Y. Shalaginov, Hong Tang, Hang Li, Li~Zhou, Jun Ding, Anuradha~Murthy Agarwal, Clara Rivero-Baleine, Kathleen~A. Richardson, Tian Gu, Juejun Hu, and Hualiang Zhang.
\newblock A deep learning approach for objective-driven all-dielectric metasurface design.
\newblock {\em ACS Photonics}, 6(12):3196--3207, 2019.

\bibitem{unet}
Olaf Ronneberger, Philipp Fischer, and Thomas Brox.
\newblock U-net: Convolutional networks for biomedical image segmentation, 2015.

\bibitem{adamw}
Ilya Loshchilov and Frank Hutter.
\newblock Decoupled weight decay regularization.
\newblock In {\em International Conference on Learning Representations}, 2019.

\bibitem{schedulefree}
Aaron Defazio, Xingyu Yang, Harsh Mehta, Konstantin Mishchenko, Ahmed Khaled, and Ashok Cutkosky.
\newblock The road less scheduled, 2024.

\bibitem{adam}
Diederik~P Kingma.
\newblock Adam: A method for stochastic optimization.
\newblock {\em arXiv preprint arXiv:1412.6980}, 2014.

\end{thebibliography}
\end{document}